\documentclass[a4paper,fleqn,usenatbib]{mnras}
\usepackage{newtxtext,newtxmath}
\usepackage[T1]{fontenc}
\usepackage{ae,aecompl}

\usepackage{graphicx}	
\usepackage{amsmath}	
\usepackage{amssymb}	

\newcommand{\kms}{{\mathrm{km~s^{-1}}}}

\title[Seismic analysis of SX Phoenicis]{Seismic analysis of the double-mode radial pulsator SX Phoenicis}

\author[Daszy\'nska-Daszkiewicz et al.]{ J. Daszy\'nska-Daszkiewicz$^{1}$\thanks{E-mail:daszynska@astro.uni.wroc.pl},
 A. A. Pamyatnykh$^{2}$, P. Walczak$^{1}$,  W. Szewczuk$^{1}$\\
$^{1}$Instytut Astronomiczny, Uniwersytet Wroc{\l}awski, Kopernika 11, 51-622 Wroc{\l}aw, Poland\\
$^{2}$Nicolaus Copernicus Astronomical Center, Bartycka 18, 00-716, Warsaw, Poland}

\date{Accepted XXX. Received YYY; in original form ZZZ}

\pubyear{2019}

\begin{document}
\label{firstpage}
\pagerange{\pageref{firstpage}--\pageref{lastpage}}
\maketitle

\begin{abstract}
	We present the results of complex seismic analysis of the prototype star SX Phoenicis. This analysis consists of a simultaneous fitting
	of the two radial-mode frequencies, the corresponding values  of the bolometric flux amplitude (the parameter $f$)
	and of the intrinsic mode amplitude $\varepsilon$.
	The effects of various parameters as well as the opacity data are examined.
	With each opacity table it is possible to find seismic models  that reproduce the two observed frequencies
	with masses allowed by evolutionary models appropriate for the observed values of the effective temperature and
	luminosity. All seismic models are in the post-main sequence phase.
	The OPAL and OP seismic models are in hydrogen shell-burning phase and the OPLIB seismic model has just finished 
	an overall contraction and starts to burn hydrogen in a shell.
	The OP and OPLIB models are less likely due to the requirement of high initial hydrogen abundance ($X_0=0.75)$ 
	and too high metallicity ($Z\approx 0.004$) as for a Population II star.\\
	The fitting of the parameter $f$, whose empirical values are derived from multi-colour photometric observations,  provides constraints
	on the efficiency of convective transport in the outer layers of the star and on the microturbulent velocity in the atmosphere.
	Our complex seismic analysis with each opacity data indicates low to moderately efficient convection in the star's envelope, described by the mixing length parameter of $\alpha_{\rm MLT}\in  (0.0,~0.7)$,
	and the microturbulent  velocity in the atmosphere of about $\xi_{\rm t}\in(4,~8)~\kms$.
\end{abstract}




\section{Introduction}

SX Phoenicis  (HD\,223065, SX Phe) is  a star with the mean visual magnitude of  $V=7.12$ mag
and a A3V spectral type. The star belongs to Population II and was discovered to be variable
by \citet{Eggen1952a, Eggen1952b}. Since then SX Phe became
a prototype for the whole class of high-amplitude and, usually, metal-poor pulsators located
inside the $\delta$ Scuti instability strip.
SX Phoenicis was a target of several studies based on photometric and spectroscopic observations.
The analysis of photometric data revealed two frequencies and their combinations
\citep[e.g., by][]{Coates1979, Rolland1991, Garrido1996} with the values $\nu_1=18.1936$ d$^{-1}$
and $\nu_2=23.3794$ d$^{-1}$ \citep{Garrido1996}.
The frequency ratio indicates that SX Phe pulsates most probably in the two radial modes;
the fundamental  one and first overtone.
These two periodicities were detected also in the radial velocity variations by \citet{Kim1993}
with the amplitudes of about 18\,km\,s$^{-1}$ for the dominant frequency and about 4\,km\,s$^{-1}$ for the secondary one.
Quite surprisingly, the recent analysis of the high-precision photometry from the TESS satellite
by \citet{Antoci2019} did not show any firm additional independent frequencies and, so far, confirmed the older results.
The values of frequencies extracted from the TESS light curve of SX Phe  are:
$\nu_1=18.193565(6) ~{\mathrm{d^{-1}}}$ and $\nu_2=23.37928(2) ~{\mathrm{d^{-1}}}$.

There is also some evidence that both pulsation periods change in a timescale of decades \citep{Landes2007}.
Moreover, for the dominant pulsational period the effective temperature varies in a huge range
from 7230 to 8210\,K and the surface gravity from 4.25 to 3.86 dex \citep{Rolland1991}.
The corresponding mean values are 7640\,K and 3.89 \citep{Kim1993}.
The recent determination of the effective temperature from spectroscopy amounts
to  $T_{\rm eff} = 7500(150)$\,K
and the luminosity derived from the Gaia DR2 data is  $\log L/L_\odot=0.844(9)$ \citep{Antoci2019}.
The metallicity of SX Phe is typical for most stars of Population II. From photometric indexes, assuming
a zero interstellar reddening, \citet{Rolland1991}  derived the value [m/H]$=-1.0$  and \citet{McNamara1997}
the value [m/H]$=-1.4$. The most recent determination in \citet{Antoci2019} amounts to $-1.00(15)$.
According to many determinations, the rotational velocity is low, $V_{\rm rot}=18(2)~\mathrm{km~s^{-1}}$,
\citep[e.g.,][]{Rodriguez2000, Antoci2019}.

There were several attempts to estimate a mass of SX Phe. \citet{Vandenberg1985} obtained $M=1.2(1)~\mathrm{M}_\odot$ using 
evolutionary tracks for the helium abundance $Y=0.25$ and metallicity $Z=0.0017$.
Other determinations were based on the two radial-mode periods and/or pulsational equation.
\citet{Dziembowski1974} indicated a small mass of about 0.2\,$\mathrm{M}_\odot$ whereas \citet{Cox1979} derived
1.1\,$\mathrm{M}_\odot$. The values around one solar mass  were obtained also by \citet{Andreasen1983}: 
$M=1.27\,\mathrm{M}_\odot$, and \citet{Eggen1989}: $M=0.91\,\mathrm{M}_\odot$.
The seismic  modelling with the early version of the OPAL opacity data \citep{Iglesias1992} was performed by \citet{Petersen1996}.
These authors showed that the period ratio of the two radial modes is best reproduced by the model with parameters:
a mass $M= 1.0M_\odot$, metallicity $Z = 0.001$, initial hydrogen abundance $X_0 = 0.70$ and age 4.07\,Gyr.
Recently, also initial results of our seismic modelling have been published in \citet{Antoci2019} and \citet{JDD2020}.

The aim of this paper is to perform seismic modelling of SX Phe in a wide space of parameters
and to study the effect of various opacity data. Besides, we try to reproduce  the bolometric flux
amplitude (the parameter $f$) which is very sensitive to physical conditions in subphotospheric layers.

In Section\,2, we present an independent mode identification from the Str\"omgren amplitudes and phases.
Then, the results of fitting the two modes are given with detailed studies of the effects of various parameters
as well as the opacity data.
An attempt to reproduce also the parameter $f$ of the two radial modes is shown in Section\,3.
The last section summarizes our results.


\section{Pulsations of SX Phoenicis}

Despite several efforts to find more periodic signals in the highly-asymmetric light curve of SX Phe, it seems that
mainly the two frequencies are responsible for the variability of the star.
From the Fourier analysis of the TESS light curve, it appeared that down to an amplitude of 1 ppt,
there are two main frequencies with the values $\nu_1=18.193565(6) ~{\mathrm{d^{-1}}}$ and $\nu_2=23.37928(2) ~{\mathrm{d^{-1}}}$ 
and the amplitudes of about 136 ppt and 33 ppt, respectively \citep{Antoci2019}.
Other five independent frequencies with very low amplitudes, but above 1 ppt, are in the frequency range (17, 50) d$^{-1}$.
Besides, many combinations and harmonic frequencies,  up to the 7th one, were identified.
In total, 27 frequency peaks were extracted from the TESS light curve.

As noticed in many earlier papers, the frequency ratio $\nu_1/\nu_2=0.778192(6)$ indicates that
these two frequencies can correspond to the consecutive radial modes: fundamental and first overtone.
Here, we add for the first time an independent mode identification based on the photometric amplitudes and phases.

\subsection{Independent mode identification}
In the framework of linear theory of stellar pulsations, assuming the zero-rotation approach,
the complex amplitude of a given mode in passband $\lambda$ is given by (e.g., \citet{JDD2002}):
$${\cal A}_{\lambda}(i) = -1.086 \varepsilon Y_{\ell}^m(i,0) b_{\ell}^{\lambda}
(D_{1,\ell}^{\lambda}f+D_{2,\ell}+D_{3,\ell}^{\lambda})\eqno(1)$$
where
$$D_{1,\ell}^{\lambda} = \frac14  \frac{\partial \log ( {\cal F}_\lambda |b_{\ell}^{\lambda}| ) } {\partial\log T_{\rm{eff}}} ,
\eqno(2a)$$
$$D_{2,\ell} = (2+\ell )(1-\ell ), \eqno(2b)$$
$$D_{3,\ell}^{\lambda}= -\left( 2+ \frac{\omega^2 R^3}{GM} \right)
\frac{\partial \log ( {\cal F}_\lambda |b_{\ell}^{\lambda}|	) }{\partial\log g}.\eqno(2c)$$
$\varepsilon$ is the intrinsic mode amplitude, $i$ is the
inclination angle and $Y_{\ell}^m$ denotes  the spherical harmonic with the degree $\ell$ and
the azimuthal order $m$. Symbols
$G,M,R,\omega$ have their usual meanings. The values of the amplitudes and phases themselves
are given by $A_{\lambda}=|{\cal A}_\lambda|$ and $\varphi_{\lambda}=arg({\cal A}_\lambda)$, respectively.

As can be concluded from the above formula, the terms $D_{1,\ell}^\lambda$, $D_{2,\ell}$
and  $D_{3,\ell}^\lambda$ correspond to the effects of pulsational changes of  temperature, geometry
and  pressure, respectively.
The term $b_{\ell}^{\lambda}$ is the integral of limb darkening weighted by the Legendre polynomial
with the $\ell$ degree. It describes the effect of disc averaging with increasing values of $\ell$.
Derivatives of the monochromatic flux,
${\cal F}_\lambda(T_{\rm eff},\log g)$, as well as limb darkening and its derivatives are calculated from static
atmosphere models. In general, their values depend on the metallicity [m/H] and microturbulence velocity $\xi_t$. 
Here, we rely on the Vienna atmosphere models (NEMO2003), which were computed with the modified versions
of the ATLAS9 code \citep{Heiter2002}, in order to include turbulent convection treatment from \citet{Canuto1996}. 
We added also our computations of model atmospheres for the microturbulent velocity $\xi_t=10$\,km\,s$^{-1}$. Since we need only fluxes and specific intensities, we calculated these quantities  
using the atmosphere model from the original NEMO grid 
and the the SYNSPEC code with the microturbulent velocity set to $\xi_t=10\,$\,km\,s$^{-1}$
\citep[e.g.,][]{2011ascl.soft09022H, 2017arXiv170601859H}.
Moreover, we derived the limb-darkening coefficients for the nonlinear formula of \citet{Claret2000} for all values of $\xi_t$.

The parameter $f$ describes the ratio of the bolometric flux perturbation to the radial displacement
for a given pulsational mode:
$$\frac{ \delta {\cal F}_{\rm bol} } { {\cal F}_{\rm bol} }=
{\rm Re}\{ \varepsilon f Y_\ell^m(\theta,\varphi) {\rm e}^{-{\rm i} \omega t} \}.\eqno(3)$$
%
The value of $f$ is complex and can be obtained from nonadiabatic computations of stellar pulsations.

To identify pulsational modes of SX Phe, we followed the method of \citet{JDD2003}.
In this method, the mode degree $\ell$, the parameter $f$ and the intrinsic mode amplitude, 
multiplied by the inclination-dependent factor, $\varepsilon Y_{\ell}^m(i,0)$ are determined simultaneously. It is achieved
by fitting the theoretical values of the photometric amplitudes and phases to their observed counterparts.
In this way, firstly, one avoids the uncertainties in the theoretical values of the parameter $f$, and secondly,
valuable constraints on the parameters of model and theory can be  derived from a comparison of the theoretical and empirical values of $f$. In fact, we derive not the pure observational values $f$ but the semi-empirical ones
because one has to adopt some atmosphere models to compute the flux derivatives and limb darkening.

The goodness of the fit can be written as
$$\chi^2=\frac1{2N-M} \sum_{i=1}^N  \frac{ |{\cal A}^o_{\lambda_i} - {\cal A}^t_{\lambda_i}|^2 }{ |\sigma_{\lambda_i}|^2}, \eqno(4)$$
where $N$ is the number of passbands $\lambda_i$ and $M$ is the number of parameters to be determined.
The methods yields two complex parameters, $\varepsilon$ and $f$, thus $M=4$.
The  symbols ${\cal A}^o$ and ${\cal A}^t$ are complex observational and theoretical amplitudes,
respectively, and $\sigma_{\lambda_i}$ are their observational errors.

\begin{table}
\centering
\caption{Amplitudes and phases in the TESS passbands (in units ppt) and the four  Str\"omgren passbands (in units mag) for the two frequencies of SX Phoenicis. The frequencies are derived from the TESS light curve \citep{Antoci2019}. }
\begin{tabular}{ccccccc}
\hline
     & $A$ [ppt], [mag] & $\varphi$ [rad]  \\
\hline
\multicolumn{3}{c}{$\nu_1= 18.193565$ d$^{-1}$} \\
\hline
TESS & 136.284(40)  &  \\
  $u$  &   0.2046(6) &  1.991(3) \\
  $v$  &   0.2786(6) &  1.864(2)\\
  $b$  &   0.2511(6) &  1.865(2)\\
  $y$  &   0.2059(5) &  1.851(2)\\
\hline
\multicolumn{3}{c}{$\nu_2=  23.379283$ d$^{-1}$} \\
\hline
TESS &     33.079(40) [ppt] &   \\
  $u$  &   0.0793(6) &  3.768(7)\\
  $v$  &   0.0993(6) &  3.647(6)\\
  $b$  &   0.0895(6) &  3.653(6)\\
  $y$  &   0.0742(5) &  3.652(7)\\
\hline
\end{tabular}
\end{table}

In the case of $\delta$ Sct and SX Phe stars, the parameter $f$ is very sensitive to the efficiency of
convective transport in the outer layers.
First successful applications of this method were demonstrated for $\delta$ Sct  pulsators by
\citet{JDD2003,JDD2005a}
and for $\beta$ Cep pulsators by \citet{JDD2005b}. In the case of B-type pulsator
valuable constraints on stellar opacities were obtained \citep[see][]{JDD2017, Walczak2019}.
Moreover, the effect of atmosphere models was shown by \citet{JDD2007}.

\begin{figure*}
	\includegraphics[width=175mm,clip]{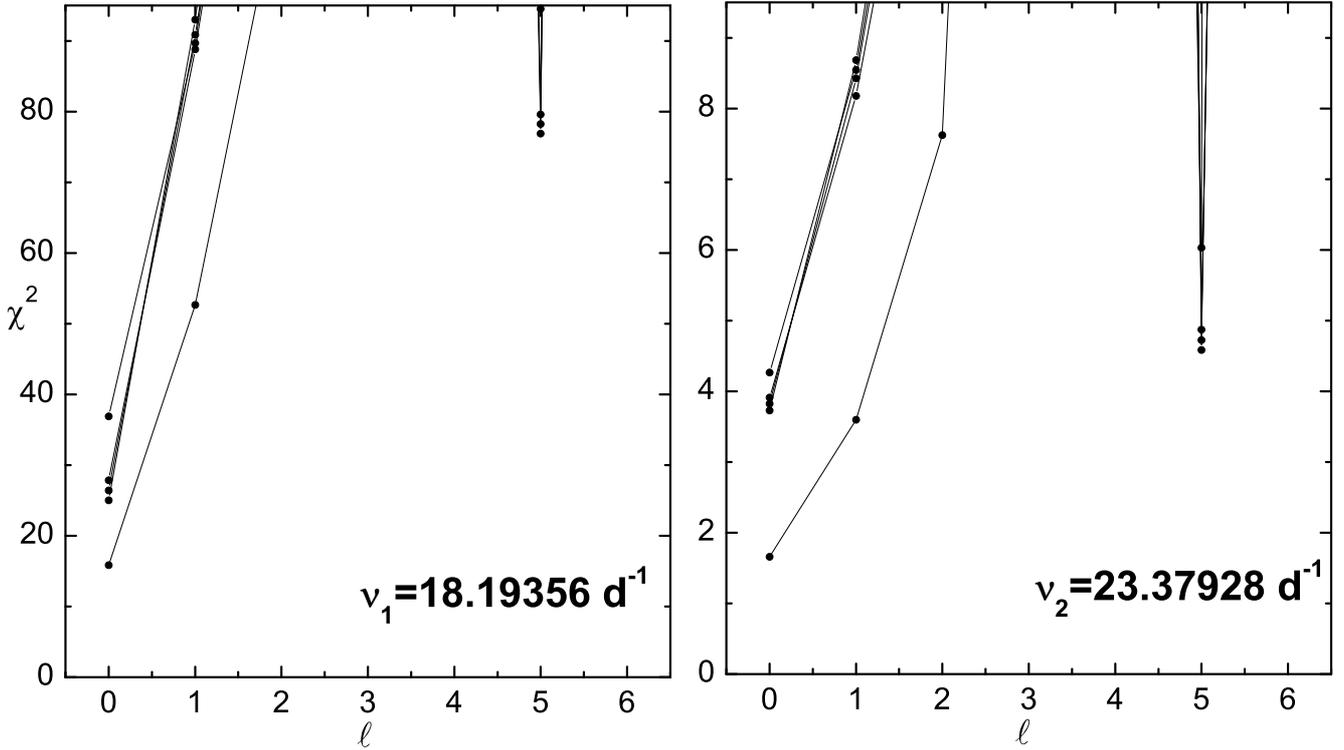}
	\caption{The values of $\chi^2$ as a function of $\ell$ for the two frequencies of SX Phe.
		NEMO model atmospheres were adopted with the metallicity [m/H]$=-1.0$
		and the microturbulent velocity $\xi_{\rm t}=2$ km$\,$s$^{-1}$. The logarithmic values
		of the effective temperature $\log T_{\rm eff}$ and the surface gravity $\log g$
		are from the range (3.855, 3.895) and (4.00, 4.15), respectively. The left panel corresponds
		to the dominant frequency peak and the right one to the secondary peak.}
	\label{fig1}
\end{figure*}

The multicolour time-series photometry of SX Phe in the four Str\"omgren passbands was performed by \citet{Rolland1991}.
The values of amplitudes and phases in the Str\"omgren and TESS bands are given in Table\,1.
In Fig.\,1,  we  show the values of the discriminant $\chi^2$ as a function of the mode degree $\ell$
for the dominant frequency (the left panel) and for the second frequency (the right panel).
Four pairs $(\log T_{\rm eff},~\log g)$, spanning the ranges (3.855, 3.895) and (4.00, 4.15), respectively, were considered.
The adopted atmospheric metallicity was [m/H]$=-1.0$.
As one can see, the radial modes for both frequencies of SX Phe are clearly preferred.

\subsection{Fitting the two radial-mode frequencies}

Evolutionary computations were performed using  the Warsaw-New Jersey code \citep[e.g.,][]{Pamyatnykh1998, Pamyatnykh1999}.
Three opacity tables were used: OPAL \citep{Iglesias1996}, OP \citep{Seaton1996, Seaton2005} and OPLIB \citep{Colgan2015, Colgan2016}.
In each case, the lower temperature range, i.e., for $\log T<3.95$, was supplemented with the data of \citet{Ferguson2005}.
The solar chemical mixture was adopted from \citet{Asplund2009}.
In all calculations the OPAL2005 equation of state was used \citep{Rogers1996, Rogers2002}.
The Warsaw-New Jersey code takes into account the mean effect of the centrifugal force, assuming solid-body rotation and constant global angular momentum during evolution.
The treatment of convection in the stellar envelope relies on the standard mixing-length theory.

Linear nonadiabatic oscillations were computed with the code of \citet{Dziembowski1977a}.
The code takes into account the effects of rotation up to the second order.
Convection flux is assumed to be constant during the pulsational cycle. This is so called the convective flux freezing approximation
which is quite good if convection is not very efficient.

We search a wide space of parameters appropriate for SX Phe.  As for the effective temperature
we allowed for the whole range published in the literature, i.e., $\log T_{\rm eff}\in (3.8483,~3.9325)$
The value of luminosity $\log L/L_{\odot}=0.844(9)$ was adopted after \citet{Antoci2019}
who determined it from the Gaia parallax using $T_{\rm eff}$ derived from the IRFM method.
However, we allowed for the $3\sigma$ error in $\log L/L_{\odot}$.
The most recent determination of metallicity gives [m/H]$=-1.00(15)$ \citep{Antoci2019}.
Depending on the solar metallicity in model atmospheres used by the authors, the values of [m/H]
translates into $Z\in(0.0010, 0.0020)$ if the solar metallicity is $Z_\odot =0.014$ and into $Z\in(0.0012, 0.0024)$ if $Z_\odot =0.017$.
The metallicity we searched encompasses much wider range, i.e,  $Z\in(0.0008, 0.0040)$.

\begin{figure}
\includegraphics[width=\columnwidth,clip]{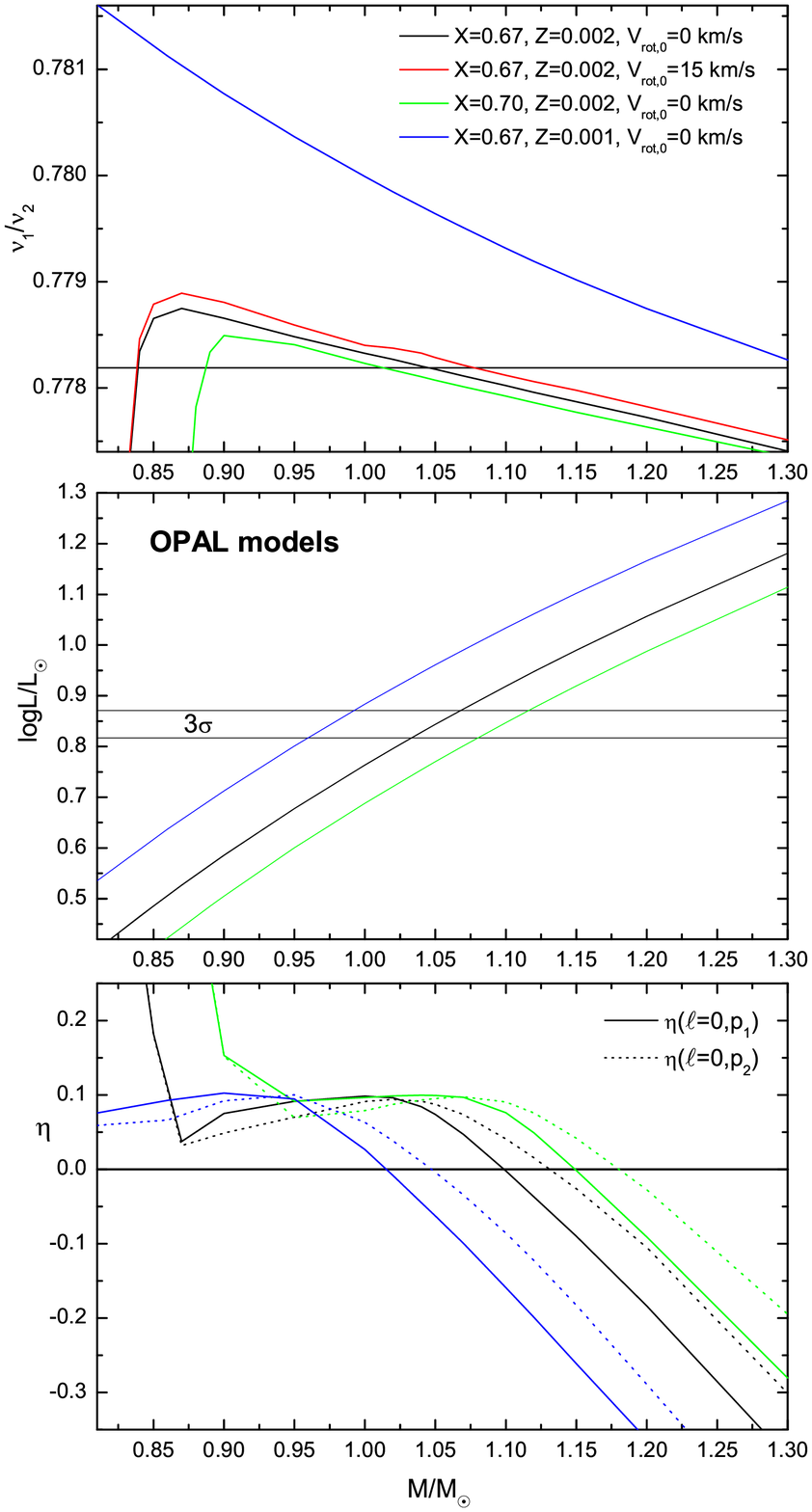}
\caption{The top panel: the frequency ratio of the radial fundamental mode to the first overtone, $\nu_1/\nu_2$,
 as a function of a mass for models that fit the observed frequency $\nu_1=18.19356$ d$^{-1}$.
 The models were computed with the OPAL opacity data. There is shown the effect of the initial hydrogen abundance ($X_0=0.67$ vs 0.70),
 metallicity ($Z=0.002$ vs 0.001) and rotation ($V_{\rm rot,0}=0$ vs $15~\mathrm{km~s^{-1}}$). The middle panel: the corresponding values of luminosity.
 The luminosities for non-rotating and rotating models overlap. The horizontal lines indicate the observed range of luminosity, as determined from the Gaia parallax,
 allowing for the $3\sigma$ error. The bottom panel: the corresponding values of the normalized instability parameter $\eta$.
 The solid lines correspond to the radial fundamental mode and the dashed ones to the first radial overtone. Modes with
 $\eta>0.0$ are excited.}
\label{fig2}
\end{figure}
\begin{table}
	\centering
	\caption{The effect of various parameters on the frequency ratio of the radial fundamental mode to the first overtone for models
		appropriated for SX Phoenicis.}
	\begin{tabular}{ccccccc}
		\hline
		\hline
		The effect of $X$  \\
		\hline
		$X$ $\nearrow$ ~~~$\Rightarrow$ ~~~ $\frac{\nu_1}{\nu_2}$  $\searrow$\\
		\hline
		\hline
		The effect of $Z$  \\
		\hline
		$Z$ $\nearrow$ ~~~$\Rightarrow$ ~~~ $\frac{\nu_1}{\nu_2}$  $\searrow$\\
		\hline
		\hline
		The effect of $V_{\rm rot}$  \\
		\hline
		$V_{\rm rot}$ $\nearrow$ ~~~$\Rightarrow$ ~~~ $\frac{\nu_1}{\nu_2}$  $\nearrow$\\
		\hline
		\hline
		The effect of $\alpha_{\rm ov}$  \\
		\hline
		$\alpha_{\rm ov}$ $\nearrow$ ~~~$\Rightarrow$ ~~~ $\frac{\nu_1}{\nu_2}$  $\searrow$\\
		\hline
		The effect of $\alpha_{\rm MLT}$  \\
		\hline
		$\alpha_{\rm MLT}$ $\nearrow$ ~~~$\Rightarrow$ ~~~ $\frac{\nu_1}{\nu_2}\approx {\rm const}$  \\
		\hline
		\hline
	\end{tabular}
\end{table}

We started the modelling with the OPAL opacity tables. For nonadiabatic convection in the outer layers we adopted 
the value of the mixing length parameter $\alpha_{\rm MLT}=1.0$. The value $\alpha_{\rm MLT}=0.0$ means that 
convective transport does not take place in the stellar envelope.
In the top panel of Fig.\,2, we plot the frequency ratio of the radial fundamental mode to the first overtone as a function of a mass.
All models reproduce exactly the dominant frequency $\nu_1=18.193565$\,d$^{-1}$ corresponding to the radial fundamental mode.
The theoretical value of the first overtone is in the range of about (23.2, 23.6)\,d$^{-1}$.
Four cases are plotted to show the effect of the initial hydrogen abundance, $X_0=0.67~vs~0.70$, metallicity,
$Z=0.002~ vs ~0.001$,  and the initial rotation $V_{\rm rot,0}=0~vs ~15~\mathrm{km~s^{-1}}$ .
The observed value of the frequency ratio is marked as a horizontal line. The corresponding values of luminosity are depicted
in the middle panel of Fig.\,2. Because the considered value of $V_{\rm rot}$ is small, the luminosities for models with and without rotation overlap.
The observed values of $\log L/L_{\odot}$ are marked with the horizontal lines, allowing for the $3\sigma$ error, i.e.,
from 0.817 to 0.871.

The first conclusion is that $\nu_1/\nu_2(M)$ is not a monotonically decreasing function of a mass.
As a consequence, there are two intersections with the line $\nu_1/\nu_2=0.778192$, corresponding to the observed value.
The first intersection is for the mass of about $M=1.05-1.07~\mathrm{M}_{\odot}$
and the second for  $M\approx 0.85-0.87~\mathrm{M}_{\odot}$.
In all cases these low mass models have much too low luminosities.
For lower metallicity $Z=0.001$ and $X_0=0.67$  the first intersection is for  much higher mass 
$M\approx 1.31~\mathrm{M}_{\odot}$  and luminosity  $\log L/L_{\odot}\approx 1.3$, and the second one for much lower 
mass $M< 0.8~\mathrm{M}_{\odot}$  (not shown in the figure). Therefore we will not consider further this case. 
\begin{figure*}
	\centering
	\includegraphics[width=\textwidth,clip]{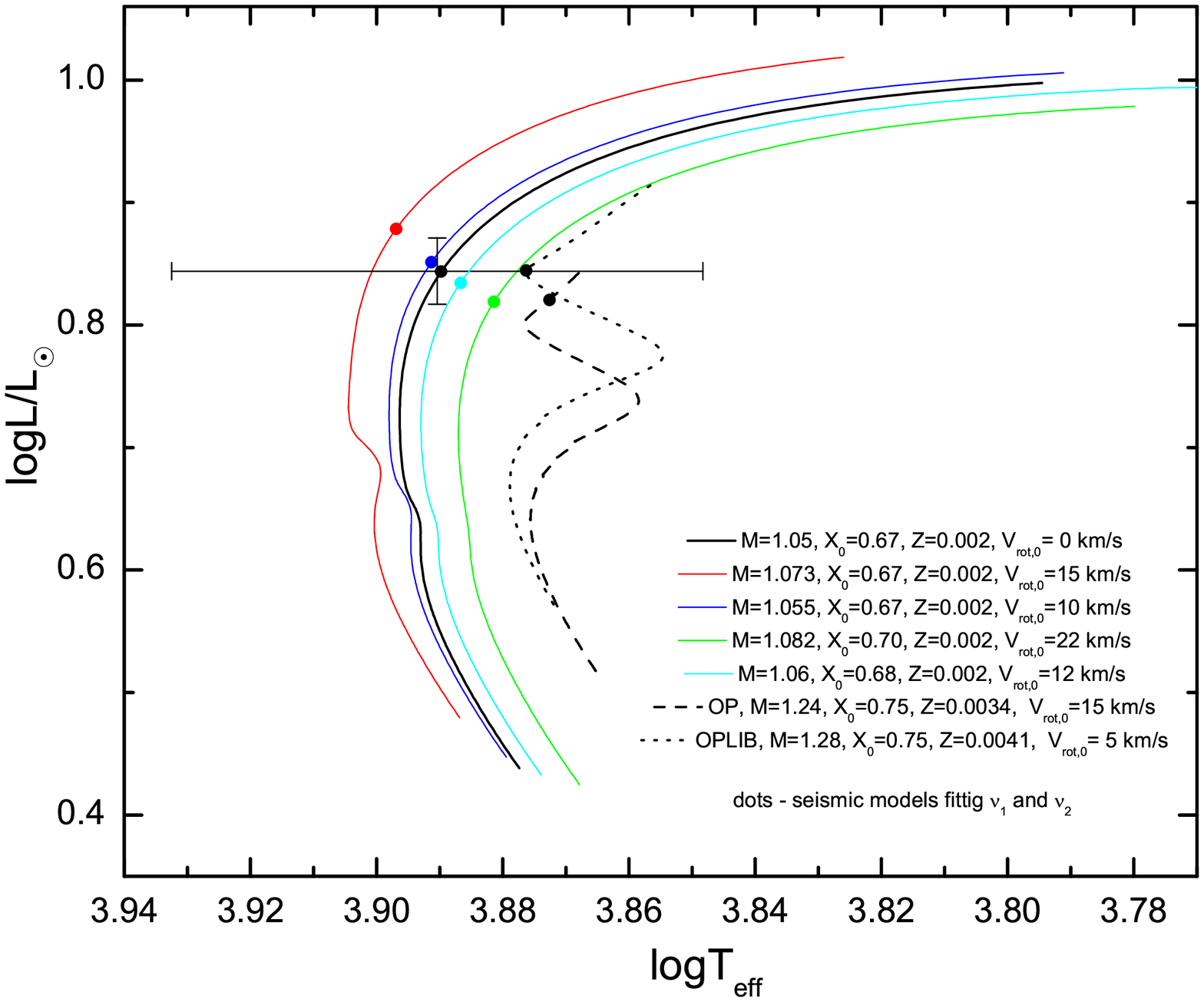}
	\caption{The HR diagram with the position of SX Phe and evolutionary tracks for various masses computed with the three opacity tables: OPAL, OP and OPLIB.
     The tracks computed adopting the OPAL data are depicted with the solid line whereas those computed adopting the OP and OPLIB data with dashed 
     and dotted line, respectively.	The models that reproduce the observed frequencies of the two radial modes are marked with dots. The 3$\sigma$ error 
     was adopted for the the GAIA luminosity. For the effective temperature the whole range available in the literature was included. Masses in the legend 
     are in solar units.}
	\label{fig3}
\end{figure*}

As one can see from the top and middle panel of Fig.\,2, only models computed with $X_0=0.67$ and $Z=0.002$ with masses
of about $M=1.05-1.07 M_\odot$ are able to reproduce both the frequency ratio and luminosity.
The model with $M=1.05 M_\odot$,  zero rotation and the parameters: $\log T_{\rm eff}=3.8897$, $\log L/L_{\odot}=0.844$,
$R=1.47 R_\odot$ has the frequency ratio $\nu_1/\nu_2=0.77818$.
The age of this model is about 2.84 Gyr, thus it is much younger than the one found by  \citet{Petersen1996}.
This model ideally reproduces both the observed frequencies and parameters, but the star must rotate.

Including the initial rotation of $V_{\rm rot,0}=15~\mathrm{km~s^{-1}}$ for the same  chemical composition we got the seismic model with the parameters: $M=1.073 M_\odot$, $\log T_{\rm eff}=3.8969$, $\log L/L_{\odot}=0.878$, $R=1.48 R_\odot$
and $\nu_1/\nu_2=0.77819$. Thus, the model has the luminosity slightly above the $3\sigma$ error of the observed value
of $\log L/L_{\odot}$.
From the top panel of Fig.\,2, we can see that decreasing the rotational velocity will shift the line $\nu_1/\nu_2(M)$
down. We found that the model with a mass $M=1.055 \mathrm{M}_\odot$ and $V_{\rm rot,0}=10~\mathrm{km~s^{-1}}$
would reproduce, both, the two frequencies and stellar parameters. The effective temperature, luminosity and radius
of this model are $\log T_{\rm eff}=3.8913$, $\log L/L_{\odot}=0.851$, $R=1.47 R_\odot$ , respectively.
The frequency ratio of the two first radial modes is $\nu_1/\nu_2=0.77819$. The current rotation is  
$V_{\rm rot}=9.8~\mathrm{km~s^{-1}}$ and the age is 2.80 Gyr.
The second intersection of the line $\nu_1/\nu_2(M)$ with the observed  frequency ratio occurs for the mass $M=0.838~M_{\odot}$ but the luminosity amounts only to about $\log L/L_{\odot}=0.46$.

For hydrogen abundance $X_0=0.70$, metallicity $Z=0.002$ and zero-rotation, the model with the mass
$M\approx 1.02~\mathrm{M}_{\odot}$  reproduces the two observed frequencies, but its luminosity is 
only $\log L/L_{\odot}=0.721$.
Increasing the initial rotational velocity to 20\,$\mathrm{km~s^{-1}}$, we got the fit of the frequencies at the mass
$M= 1.082~\mathrm{M}_{\odot}$, effective temperature $\log T_{\rm eff}=3.8814$ and the luminosity $\log L/L_{\odot}= 0.819$, which is at the edge of the $3\sigma$ error. The model is 3.07 Gyr old.
The rotation of this model is $V_{\rm rot}=23.6~\mathrm{km~s^{-1}}$ and its radius $R=1.48 R_\odot$.
One more best seismic model from our search, which reproduces both the observed frequencies and luminosity of SX Phe, has the following parameters:
$X_0=0.68$, $Z=0.002$, $M=1.06~\mathrm{M}_\odot$, $\log T_{\rm eff}=3.8867$, $\log L/L_{\odot}=0.834$, $R=1.47 R_\odot$, and the current
rotation $V_{\rm rot}=14.4~\mathrm{km~s^{-1}}$. The age of this model is about 2.92 Gyr.

The five seismic models described above, with corresponding evolutionary tracks, are depicted with dots in Fig.\,3.
All of them are in the post-main sequence phase and burn hydrogen in the shell.
There are shown also the OP and OPLIB seismic models with corresponding evolutionary tracks.
They will be discussed in the next subsection.

The summary of the studied effects on the frequency ratio in the considered range of parameters is given in Table\,2.
The effect of overshooting from the convective core in the main sequence phase on the value of $\nu_1/\nu_2$ 
for models with $M\gtrsim 1.1~\mathrm{M}_{\odot}$   is negligible and we did not include
it in our computations. For example, the difference in $\nu_1 /\nu_2$ between the models with the overshooting
parameter $\alpha_{\rm ov}=0.0$ and $\alpha_{\rm ov}=0.2$ is of the order of $10^{-5}$. Thus it is at the level of
the numerical accuracy.
Similarly, the value of the MLT parameter $\alpha_{\rm MLT}$, describing the efficiency of nonadiabatic convection 
in the stellar  envelope, has small effect on the frequency ratio.
For example, the difference between the frequency ratio of models with masses around $M=1.05~\mathrm{M}_{\odot}$
computed with $\alpha_{\rm MLT}=0.0$ and  $\alpha_{\rm MLT}=1.0$ is of the order of $10^{-5}$.

It is quite probable that SX Phe itself is a blue straggler as many SX Phoenicis variables.
Blue stragglers are presumably formed by the merger of two stars
or by interactions in a binary system and, as a consequence, they may have
enhanced helium abundance \citep[e.g.,][]{McNamara2011, Nemec2017}.

Therefore, we consider models with a lower hydrogen (higher helium) abundance to be more preferred.
These models rotate with the velocity of $10-15~\mathrm{km~s^{-1}}$.
The projected rotational velocity of SX Phe is $V_{\rm rot}\sin i=18(2)~\mathrm{km~s^{-1}}$, but most probably this value is overestimated
because of a significant contribution of pulsation to the broadening of spectral lines.
On the other hand, we cannot absolutely rule out the model with $X_0=0.70$, $Z=0.002$, 
$M=1.082~\mathrm{M}_\odot$, $\log T_{\rm eff}=3.8814$,
and $\log L/L_{\odot}= 0.819$ (if we accept the $3\sigma$ error),  rotating with the speed of
$23.6~\mathrm{km~s^{-1}}$.

To fully accept the seismic model, we have to ask about excitation of the two first radial modes. In the bottom panel of Fig.\,2, we plotted
the instability parameter $\eta$ for models considered in the two upper panels. The parameter $\eta$ is a normalized work integral
and it is greater than zero for unstable (excited) pulsational modes. The solid lines represent the fundamental mode whereas the dashed ones the first overtone.
The values of $\eta$ for non-rotating and rotating models overlap. The horizontal line indicates $\eta=0.0$.
As one can see, in the case of $X_0=0.67, ~Z=0.002$, the models with masses $M<1.1~\mathrm{M}_\odot$ have 
both radial modes unstable. As for other seismic models described above, all of them have also both the radial fundamental 
as well as first overtone modes unstable.


\subsection{The effect of opacity data}

\begin{figure}
	\includegraphics[width=\columnwidth,clip]{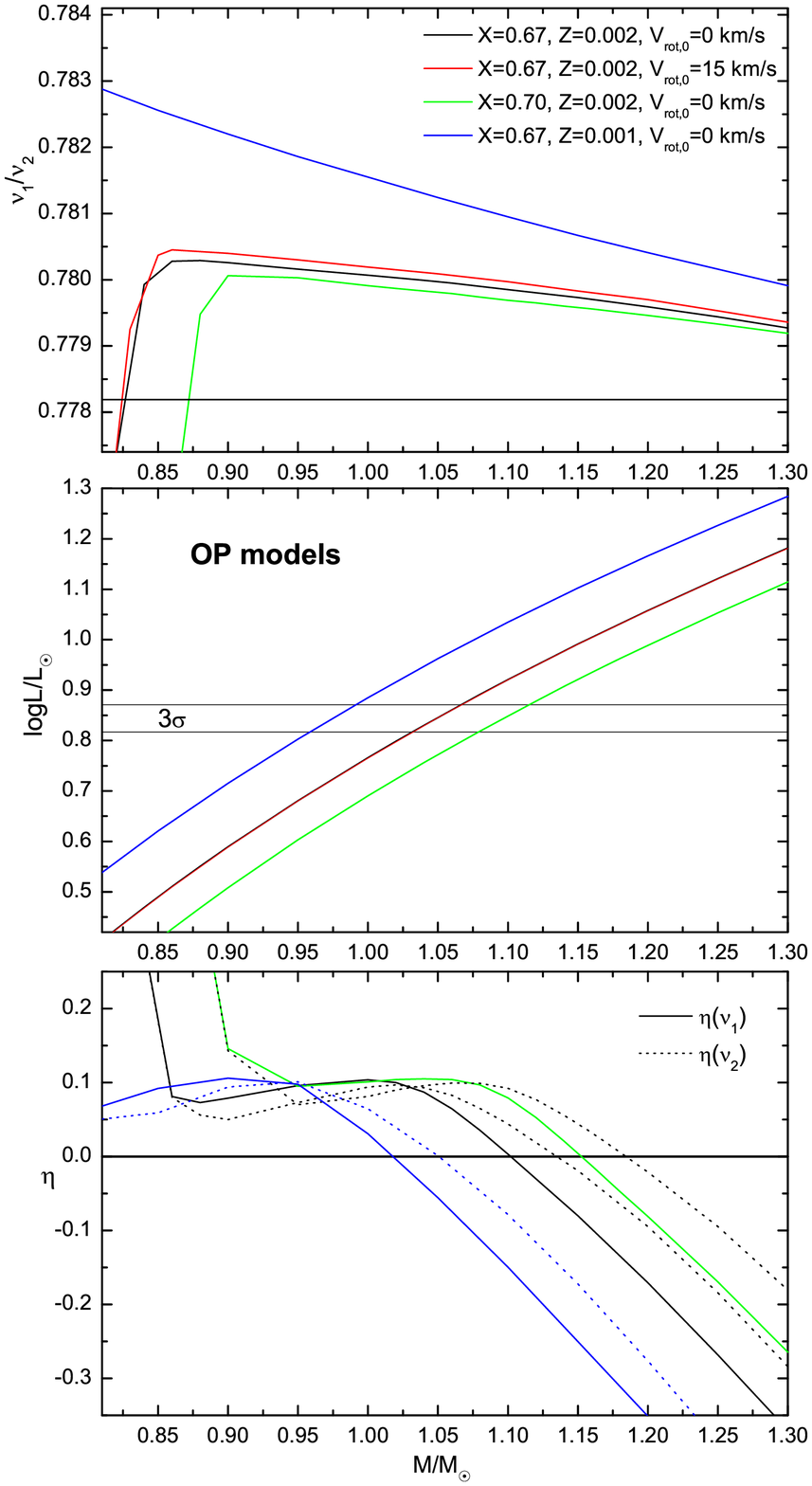}
	\caption{The same as in Fig.\,2 but evolutionary models were computed adopting the OP opacities. }
	\label{fig4}
\end{figure}
\begin{figure}
	\includegraphics[width=\columnwidth,clip]{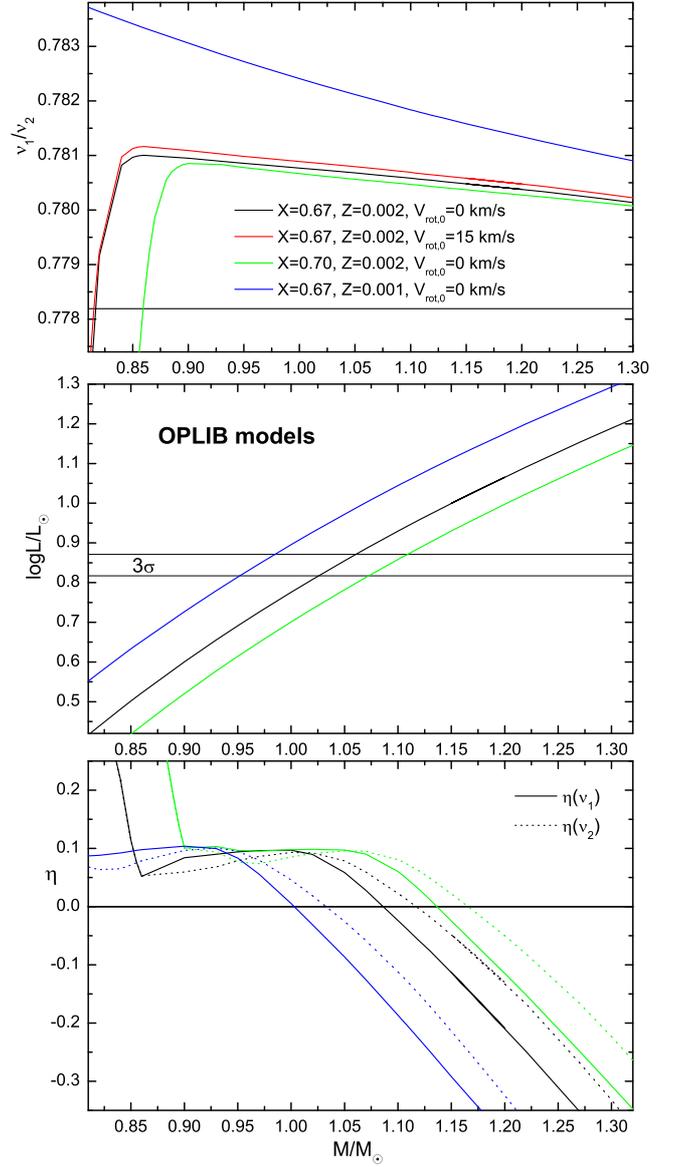}
	\caption{The same as in Fig.\,2 but evolutionary models were computed adopting the OPLIB opacities. }
	\label{fig5}
\end{figure}

In the next step, we performed the same seismic modelling using stellar opacities from the OP and OPLIB projects. The results are presented in the same way as
for the OPAL models in Figs.\,4 and 5. As one can see from the top panels, now the values of the frequency ratio $\nu_1/\nu_2$ of the models
fitting the frequency $\nu_1$ are much higher than the observed value both for the OP models (Fig.\,4) and for the OPLIB models (Fig.\,5).
Thus, in the allowed range of mass and luminosity, there is no model with the frequency ratio $\nu_1/\nu_2=0.77819$.
The intersections occurring  at $M=0.81-0.87 M_\odot$ have luminosities $\log L/L_{\odot}<0.55$. On the other hand,
the models computed with the three sources of opacity data
have very similar values of $\log L/L_{\odot}$ (cf. middle panels of Figs.\,2, 4, 5).

Considering what we have learned from seismic modelling of B-type stars \citep{JDD2017, Walczak2019},
this result is a bit surprising. 
In case of B-type pulsators with masses around $8 - 12 M_\odot$, seismic models computed with OPAL and OPLIB data had similar parameters. In the case of SX Phe, the OP and OPLIB seismic models are rather similar.
\begin{figure*}
	\centering
	\includegraphics[width=0.48\textwidth,clip]{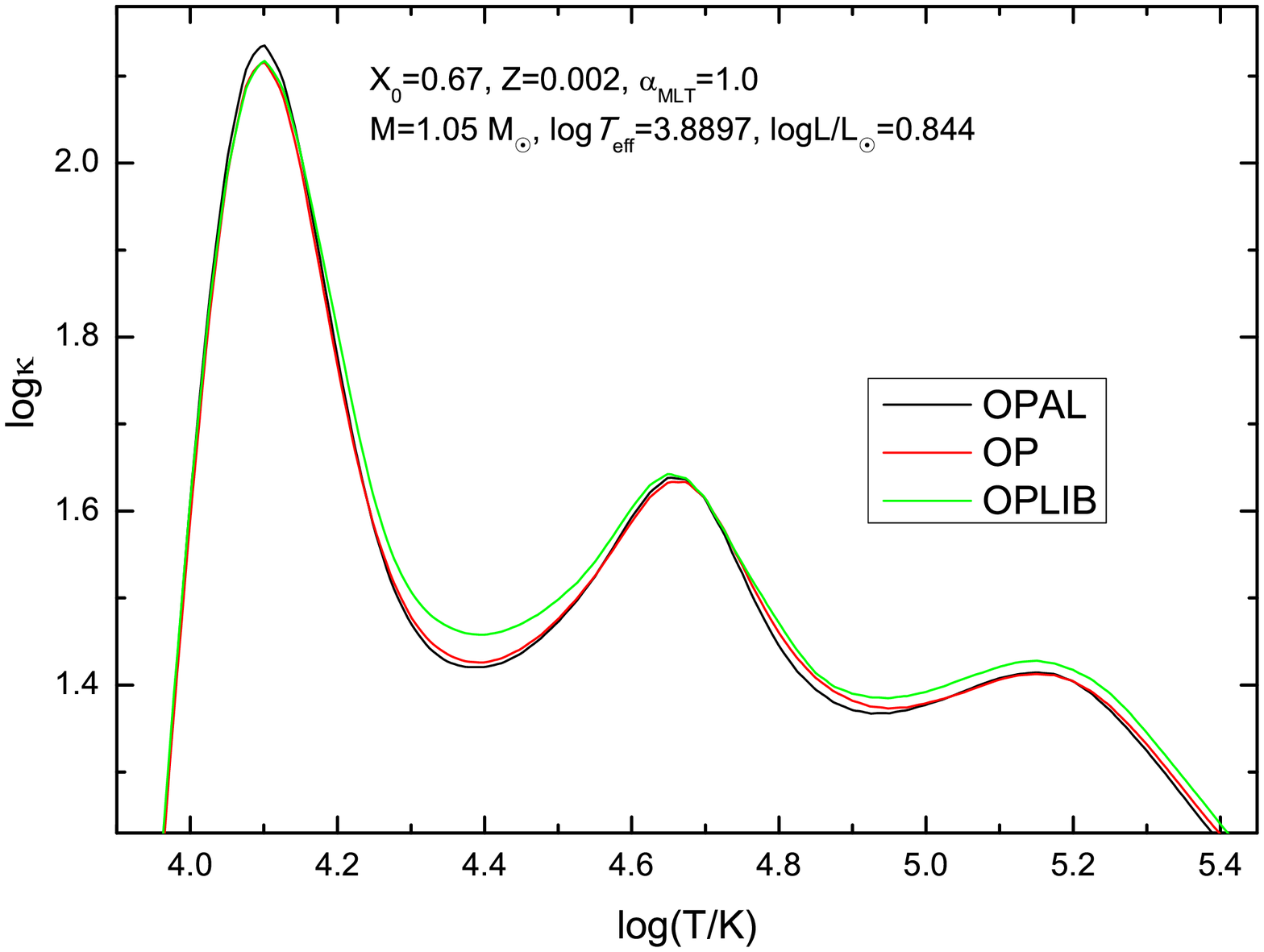}
	\includegraphics[width=0.48\textwidth,clip]{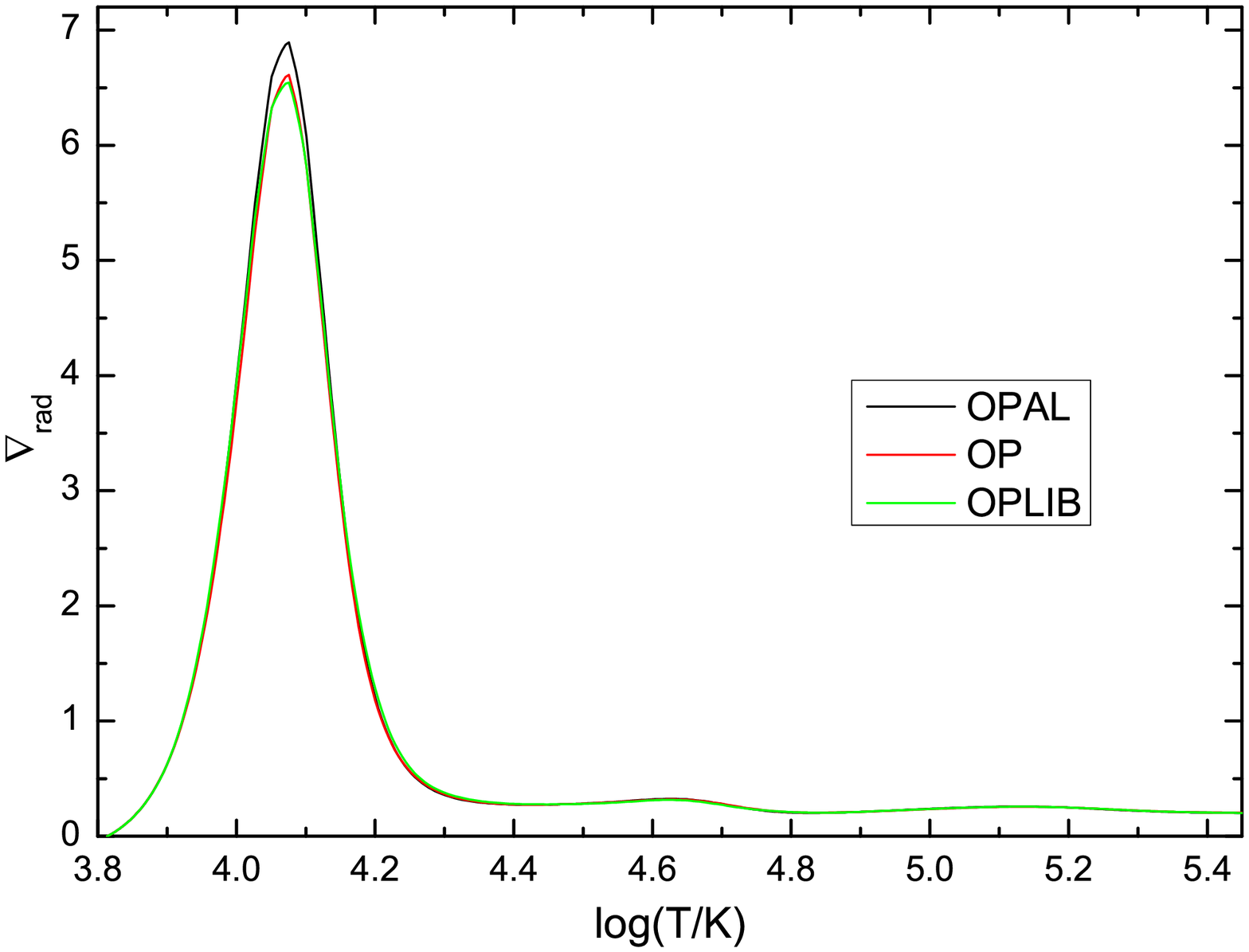}
	\caption{The left panel: the run of the mean Rosseland opacity inside the models computed for the same parameters but with different opacity data:
		OPAL, OP and OPLIB. The OPAL model reproduce the observed frequencies of the two radial modes of SX Phe as well as its effective temperature and
		GAIA luminosity. The model  parameters are given in the legend. The right panel: the corresponding values of the radiative gradient $\nabla_{\rm rad}$.}
	\label{fig6}
\end{figure*}

From the bottom panels of Figs.\,4 and 5, we can see that the instability of the two radial modes are not affected by the opacity data.
This means that in the region where pulsations are excited, i.e., around $\log T\approx 4.66$,
the mean opacity and its derivatives must be very close.
This is the case if we compare the values of $\kappa(\log T\approx 4.66)$ and its temperature derivative for the model with parameters:
$X_0=0.67$, $M=1.05~\mathrm{M}_\odot$ $\log T_{\rm eff}\approx3.89$, $\log L/L_{\odot}\approx 0.84$, $V_{\rm rot}=0$, computed with the three opacity data.
The mean opacity for this model is shown in the left panel of Fig.\,6. The corresponding values of the frequency ratio are:
$\nu_1/\nu_2({\rm OPAL})=0.77818$, $\nu_1/\nu_2({\rm OP})=0.77992$, $\nu_1/\nu_2({\rm OPLIB})=0.78080$. Making the same comparison for other masses, we can say that in general, for a given mass which reproduces the frequency $\nu_1$ we have:
$$\frac{\nu_1}{\nu_2}({\rm OPAL})<\frac{\nu_1}{\nu_2}({\rm OP})<\frac{\nu_1}{\nu_2}({\rm OPLIB}), \eqno(5a)$$
thus
$$\nu_2({\rm OPAL})>\nu_2({\rm OP})>\nu_2({\rm OPLIB}).\eqno(5b)$$
So, what is the reason for such differences in the frequency ratio?

From a very simple consideration for a homogeneous and adiabatic model, the frequency ratio of the radial fundamental mode
to the first overtone can be expressed as a function of the adiabatic index $\Gamma_1$
\citep[e.g.,][]{Kippenhahn2012}:
$$\left(\frac{\nu_1}{\nu_2}\right)=\frac{3\Gamma_1-4}{10\Gamma_1-4}, \eqno(6)$$
where $\Gamma_1>4/3$.
Although, the above formula is far from strict, we can assume at least such proportionality. It implies
that $\nu_1/\nu_2$ is always a monotonically increasing function of $\Gamma_1$. Hence, we can infer that
$$\Gamma_1({\rm OPAL})<\Gamma_1({\rm OP})<\Gamma_1({\rm OPLIB}), \eqno(7)$$
whereas for the radiative gradient $\nabla_{\rm rad}=(\partial\log T/\partial\log P)_{\rm rad}$ we should have
$$\nabla_{\rm rad}({\rm OPAL})>\nabla_{\rm rad}({\rm OP})>\nabla_{\rm rad}({\rm OPLIB}).\eqno(8)$$
In the right panel of Fig.\,6, we depicted the values of the radiative gradient for the OPAL, OP and OPLIB models.
As one can see, the only difference is around $\log T=4.1$ where hydrogen ionization takes place. In other regions
the values of $\nabla_{\rm rad}$ for models computed with the three opacity tables are equal up to the numerical accuracy.
Indeed, the value of $\nabla_{\rm rad}(\log T=4.1)$ for the OPAL model is the largest one.
Thus, it seems that this region is responsible for the subtle differences in the frequency ratio.

Of course, the natural question arises: is it possible to achieve the fit of the two frequencies and luminosity
with OP and OPLIB opacities
by changing, in a reasonable range, such parameters as mass, hydrogen abundance and metallicity?

The best fit with the OP data we obtained at the parameters: $M=1.24~\mathrm{M}_{\odot}$, $X_0=0.75$, $Z=0.0034$ and $V_{\rm rot}=17~\kms$.
The radius of this model is slightly larger comparing to the radii of the OPAL seismic models and amounts to $R=1.55~R_{\odot}$.
The logarithmic values of the effective temperature and luminosity are $\log T_{\rm eff}=3.8725$ and $\log L/L_{\odot}= 0.820$, respectively. The age of this model is 2.62 Gyr.
The position of the OP seismic model  and the corresponding evolutionary track are shown on the HR diagram in Fig.\,3.
As one can see, the model is after an overall contraction, in the hydrogen shell-burning phase.

Similarly, using the OPLIB opacities, we had to increase significantly the initial hydrogen abundance and metallicity,
up to $X_0=0.75$, $Z=0.0041$, respectively. The best OPLIB seismic model has the following parameters: $M=1.28~\mathrm{M}_{\odot}$,  $R=1.56~R_\odot$
$\log T_{\rm eff}=3.8763$ and $\log L/L_{\odot}= 0.845$ and rotates with the velocity of $V_{\rm rot}=5.5~\kms$.
Its age is 2.31 Gyr. The position of the OPLIB seismic model  and the corresponding evolutionary track
are also shown on the HR diagram in Fig.\,3. The model has just finished the overall contraction phase.

As one can see, the OP and OPLIB models have rather high metallicity and initial hydrogen abundance. 
Therefore, we consider these seismic models less probable but we cannot exclude them with 100 per cent certainty.

\section{Constraints from the parameter $\lowercase{f}$ and intrinsic mode amplitude $\varepsilon$}

The method of mode identification described in Sect.\,2.1, besides the mode degree $\ell$, provides
the semi-empirical values of the two complex quantities: the parameter $f$ and the mode amplitude $\varepsilon$.
The parameter $f$ is the  amplitude of the radiative flux variations at the level of the photosphere
and $\varepsilon$ give the relative radius variations. These two parameters are semi-empirical because their values
depend on the model atmospheres, in particular, on the metallicity and microturbulent velocity $\xi_t$.
However, for simplicity from now on,  we will call the determined values of $f$ and $\varepsilon$ as "empirical".
To this aim we used the time-series multicolour photometry of \citet{Rolland1991}.

As we mentioned, in Sect.\,2.1, the diagnostic potential of this method is huge as a comparison
of the theoretical and empirical values of $f$ yields valuable constraints on parameters of the model and theory.
In particular, for AF-pulsators, one can expect valuable constraints on the efficiency of convective transport in the outer layers.

\subsection{The values of $f$ and $\varepsilon$ for models fitting the dominant frequency }

In Fig.\,7, we show a comparison of the empirical and theoretical values of $f$ as a function of a mass.
All models reproduce the frequency $\nu_1=18.193565$ d$^{-1}$ corresponding to the radial fundamental mode.
The left panels  correspond to the dominant mode frequency $\nu_1=18.193565$ d$^{-1}$
and the right panels to the second frequency $\nu_2\approx 23.38$ d$^{-1}$. In the top panels, a run of the real part
of $f$ is shown ($f _R$) and in the bottom panels - the imaginary part of $f$ ($f_I$).
Theoretical models were computed for the chemical composition $X_0=0.67$, $Z=0.002$, the OPAL opacities and
the five values of the mixing length parameter $\alpha_{\rm MLT}=0.0,~0.5,~1.0,~1.5,~2.0$.
The rotation was not taken into account.
The empirical values of $f$ were derived adopting Vienna model atmospheres for the metallicity [m/H]=-1.0
and the four values of microturbulent velocity $\xi_t=2,~4,~8,~10~\mathrm{km~s^{-1}}$ .
The model which fits the two observed frequencies has a mass $M=1.05-1.06 M_{\odot}$ as was described in Sect.\,2.2.
As one can see from the left panels of Fig.\,7, for the dominant frequency, also around this mass, we got the agreement between the theoretical and empirical values of $f$, simultaneous for the real and imaginary part, if the MLT parameter is below 1.0 and the microturbulent velocity is about 8 ~$\mathrm{km~s^{-1}}$.
Worse agreement was obtained for the second frequency $\nu_2$. In that case, for $M\approx 1.05 M_{\odot}$,
one can adjust the real part $f_R$ for $\alpha_{\rm MLT}<0.5$ and $\xi_t\approx 10 ~\mathrm{km~s^{-1}}$, whereas
the imaginary part $f_I$ for $\alpha_{\rm MLT}<1.0$ and $\xi_t\in(2,~4)~\mathrm{km~s^{-1}}$.
Nevertheless, the result is promising because the model reproduces two frequencies and the value of $f$
for the dominant frequency, simultaneously.
The best fit of the theoretical and observed amplitudes and phases is achieved for the model atmospheres with $\xi_t=4 ~\mathrm{km~s^{-1}}$.
The results obtained for  $\xi_t= 10 ~\mathrm{km~s^{-1}}$ are rather excluded because the goodness of the fit, $\chi^2$,
is about 3--4 times worse that those obtained with $\xi_t=4 ~\mathrm{km~s^{-1}}$.
We will come back to the detailed comparison of the empirical and theoretical values of $f$ for seismic models in the next
subsection.
\begin{figure*}
 \centering
 \includegraphics[width=0.48\textwidth,clip]{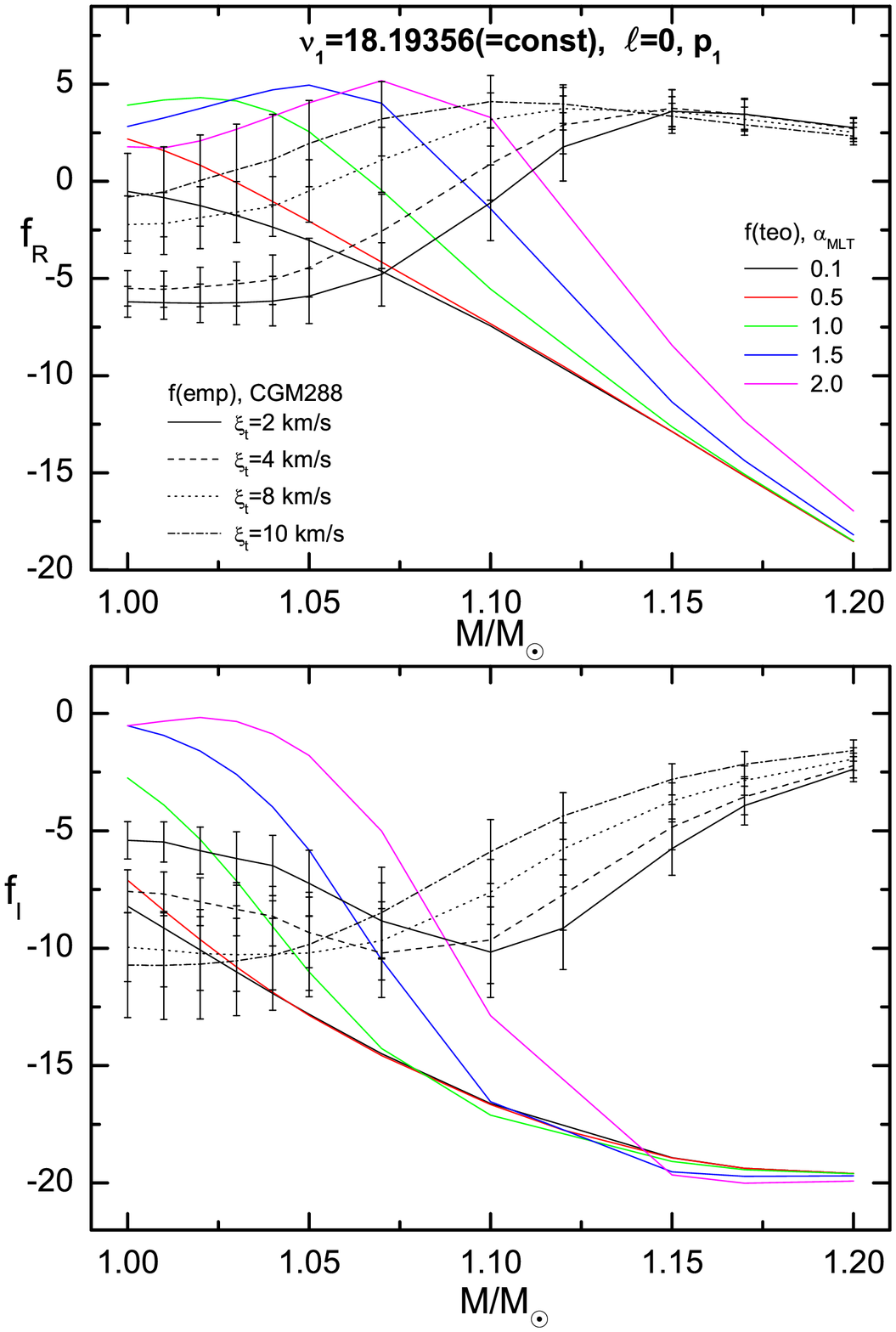}
 \includegraphics[width=0.48\textwidth,clip]{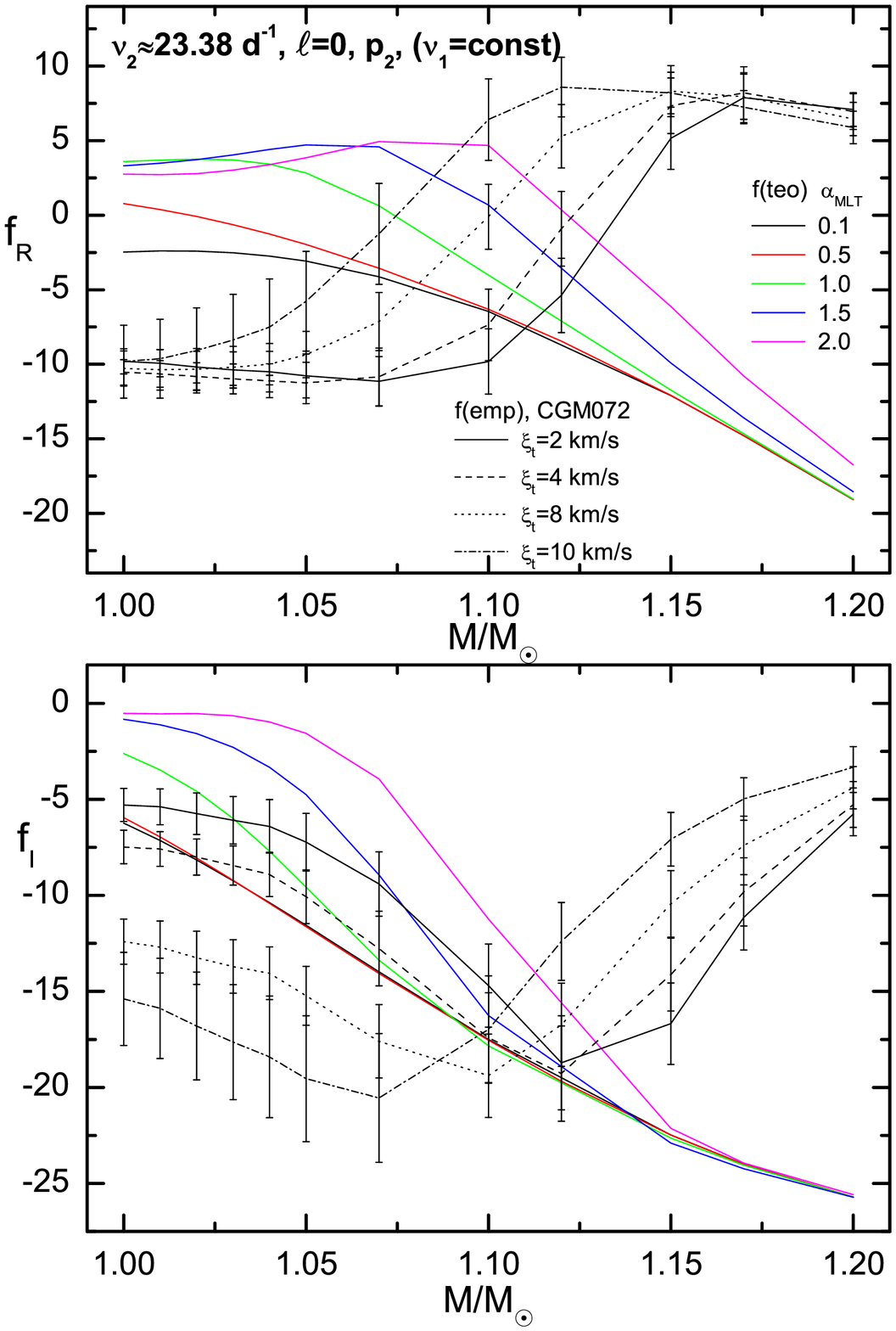}
  \caption{A comparison of the theoretical and empirical values of $f$ in a function of a mass. All models fit the radial fundamental mode $\nu_1=18.193565$ d$^{-1}$.
   The (semi)empirical values of $f$ were determined from the Str\"omgren amplitudes and phases assuming the Vienna model atmospheres with different values of the
   microturbulent velocity $\xi_t$.}
  \label{fig7}
\end{figure*}

Now let us discuss the values of the second parameter derived from the method, i.e., the intrinsic mode amplitude $\varepsilon$.
In general, the method provides the value of $\varepsilon Y_{\ell}^m(i,0)$ but in the case of radial pulsation
the absolute value of $\varepsilon$ is obtained.
The parameter $\varepsilon$ defines the local radial displacement of the surface element, i.e.,
$$\frac{\delta r(R,\theta,\varphi)}{R}=  \varepsilon Y_\ell^m (\theta,\varphi) {\rm e}^{-{\rm i}\omega t},\eqno(9)$$
where ${\rm i}$ is the imaginary unit. With the normalization of the spherical harmonics:
$$Y^m_\ell(\theta,\phi)=(-1)^{\frac{m+|m|}{2}} \sqrt{ \frac{(2\ell+1)(\ell-|m|)!}{(\ell+|m|)!} } P_{\ell}^{|m|}(\cos\theta){\rm e}^{{\rm i} m\phi},\eqno(10)$$
$|\varepsilon|$ is the r.m.s. value of $\delta r/R$ over the star surface.
On the other hand the empirical value of $\varepsilon$ can be estimated from the amplitude of radial velocity variations.
For nonradial linear pulsation, the complex formula for this amplitude is as follows \citep{Dziembowski1977b}
$$A_{\rm Vrad}(i) =  {\rm i}\varepsilon \omega R Y_\ell^m (i,0) \left(u_{\ell} + \frac{GM}{R^3\omega^2} v_{\ell}\right), \eqno(11) $$
where $u_{\ell}$  and $v_{\ell}$ are another disc averaging factors expressed by the integrals from the limb darkening and Legendre polynomials.

For the radial pulsation, $Y_\ell^m (i,0)=1$,  the factor $v_{\ell}$ is equal zero and the value of $u_{\ell}$ is 0.708 for the visual band,
adopting model atmospheres with [m/H]$=-1.0$ and $T_{\rm eff},~\log g$ appropriate for SX Phe. Thus, we get for the absolute value of $\varepsilon$
$$|\varepsilon(A_{\rm Vrad}, R)|=\frac{A_{\rm Vrad}}{0.708\omega R}.\eqno(12)$$
According to \citet{Kim1993}, the observed amplitude of the radial velocity  variations for the dominant frequency
is $18.5(5)~\kms$ and for the second frequency $4.0(5)~{\kms}$. Assuming the seismic  radius $R_{\rm s}=1.47 R_\odot$,
we get $|\varepsilon(A_{\rm Vrad}, R)|=0.0190(5)$ for $\nu_1$ and $|\varepsilon(A_{\rm Vrad}, R)|=0.0032(5)$ for $\nu_2$.

It is important to add that we cannot attach the above equation to the system of equations for the photometric amplitudes
(Eq.\,2) and derive simultaneously the parameters $\varepsilon$ and $f$ , because the time span between photometric and spectroscopic observations is too large.
The Str\"omgren photometry was gathered in 1988 \citep{Rolland1991} and the spectroscopy in 1976 \citep{Kim1993}.
Thus, the phases of the light variations and radial velocity are not consistent. Nevertheless, it still makes sense to compare
the values of the intrinsic amplitudes $\varepsilon$ derived in Sect.\,3.1 with the estimate from Eq.\,12.

In Fig.\,8, we plot the empirical values of $|\varepsilon|$ as obtained from our method for models depicted in Fig.\,7.
As before, four values of the microturbulent velocity were considered $\xi_t=2,~4,~8$ and 10 $\kms$.
The top and bottom panel corresponds to the first and second frequency, respectively. The horizontal lines mark
 the range of $|\varepsilon|$ estimated from the amplitude of the radial velocity derived by \citet{Kim1993}.
As one can see, we got the agreement of $|\varepsilon|$ between the values from our method
and the estimates from $A_{\rm Vrad}$ for both frequencies simultaneously if the mass is less than about 1.15$M_{\odot}$
and the microturbulent velocity in the atmosphere is at least 4 $\kms$.
This result is in agreement with previous constraints derived from the parameter $f$.
We can also tentatively say that the intrinsic amplitude of the radial fundamental mode of SX Phe
is about six times the intrinsic amplitude of the radial first overtone mode.

\begin{figure}
	\includegraphics[width=\columnwidth,clip]{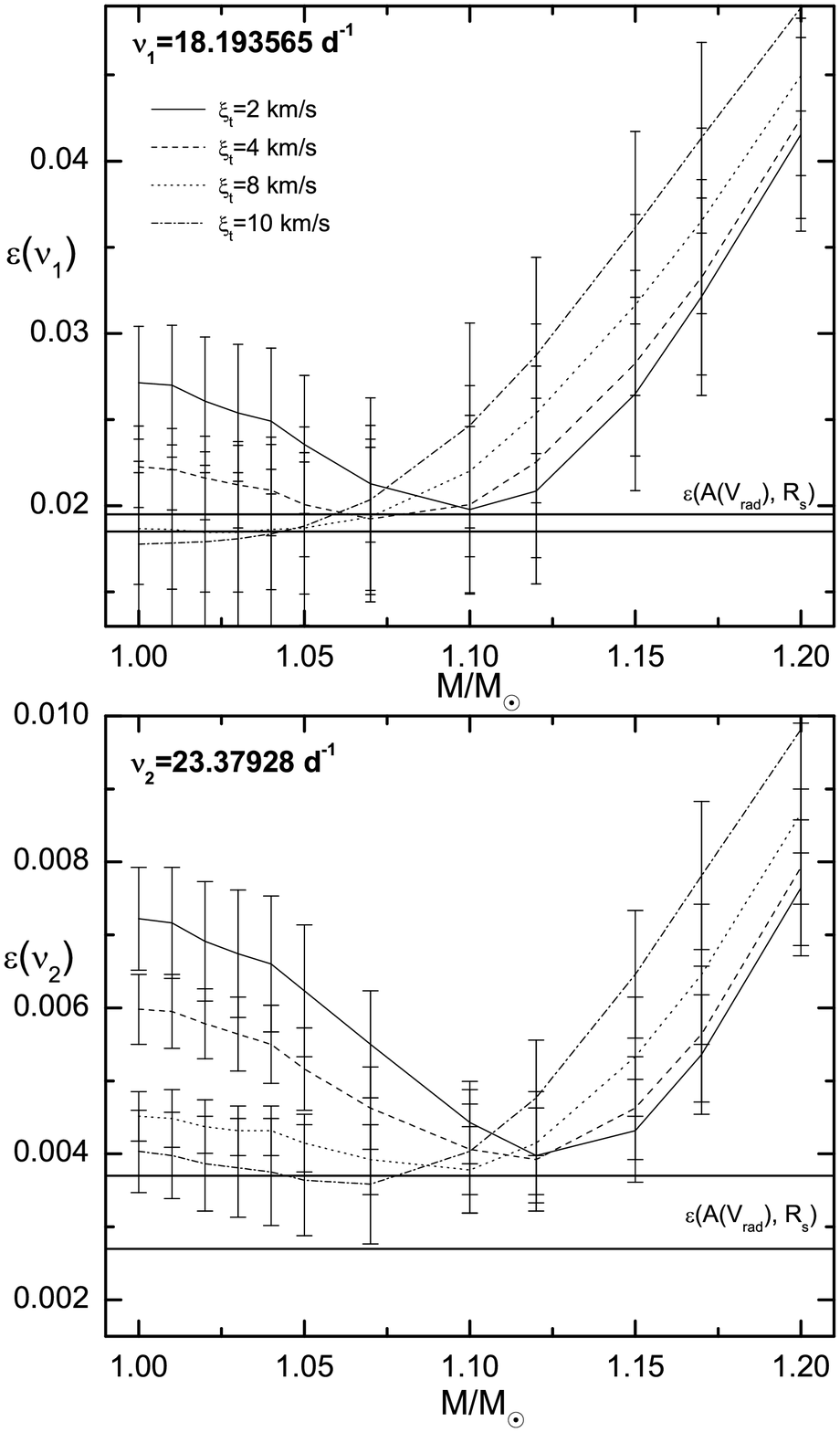}
	\caption{The empirical values of the intrinsic mode amplitude $|\varepsilon|$ for models considered in Fig.\,7. The top panel correspond to the dominant frequency (the radial fundamental mode) and the bottom panel to the second frequency
    (the first radial overtone). The horizontal lines mark the range of $|\varepsilon(A_{\rm Vrad}, R)|$ estimated from the radial velocity amplitudes
          and assuming
         the seismic radius $R_{\rm s}=1.47 R_\odot$.  }
	\label{fig8}
\end{figure}

\subsection{The values of $f$ for models fitting the two frequencies: the effect of opacities }

Now, we return to the parameter $f$ and will make detailed comparisons of their empirical and theoretical values for seismic models described in Sect.\,2. These models reproduce the two radial-mode frequencies and have the effective temperature
and luminosity within the $3\sigma$ errors of the observed values.
In Fig.\,9, we show such comparison for the OPAL seismic model with the chemical composition $X_0=0.68,~Z=0.002$ and the parameters:
$M=1.06 M_\odot,~\log T_{\rm eff}=3.8867,~\log L/L_{\odot}=0.834,~R=1.47 R_\odot$. Its rotational velocity is $14.4~\mathrm{km~s^{-1}}$.
The left panel corresponds to the dominant frequency $\nu_1=18.193565$\,d$^{-1}$
and the right panel to the second frequency $\nu_2=23.37928$\,d$^{-1}$.
As one can see, for the dominant frequency we have the agreement between the empirical and theoretical values of $f$
if the MLT parameter is  in the range (0.5,~1.0) and  $\xi_t\in (8,~10)~\mathrm{km~s^{-1}}$.
The agreement for $\xi_t=8 ~\mathrm{km~s^{-1}}$ is marginal only if $\alpha_{\rm MLT}=0.7$.
For the second frequency we could not match the empirical and theoretical values of $f$ for any microturbulent velocity.
The frequency $\nu_2$ has the photometric amplitude about 3 times smaller than the dominant frequencies. Thus, the amplitudes and phases of $\nu_2$ could be
derived not enough precisely. Certainly, new photometric time-series observations made simultaneously
with time-series spectroscopy, to include also the radial velocity into the method, could help settle this point.

For the OPAL seismic model computed with the initial hydrogen abundance $X_0=0.67$ the agreement is worse and only possible for
the microturbulent velocity $\xi_t=10 ~\mathrm{km~s^{-1}}$. However, determinations of the empirical parameter $f$ with
$\xi_t=10 ~\mathrm{km~s^{-1}}$ have much worse goodness of the fit (2$-$4 times larger $\chi^2$ in Eq.\, 4)
than for other values of $\xi_t$.

\begin{figure*}
 \centering
 \includegraphics[width=0.48\textwidth,clip]{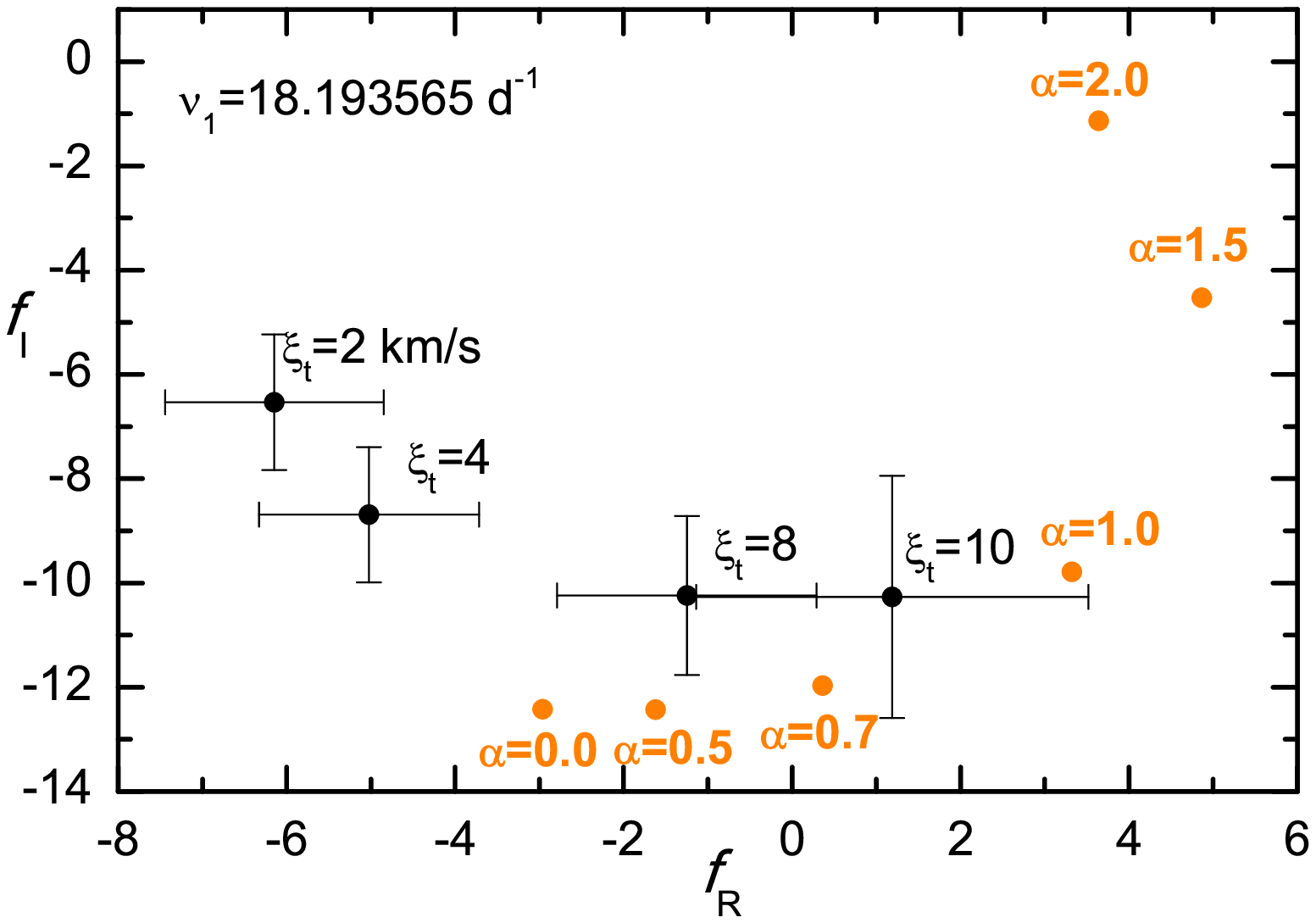}
 \includegraphics[width=0.48\textwidth,clip]{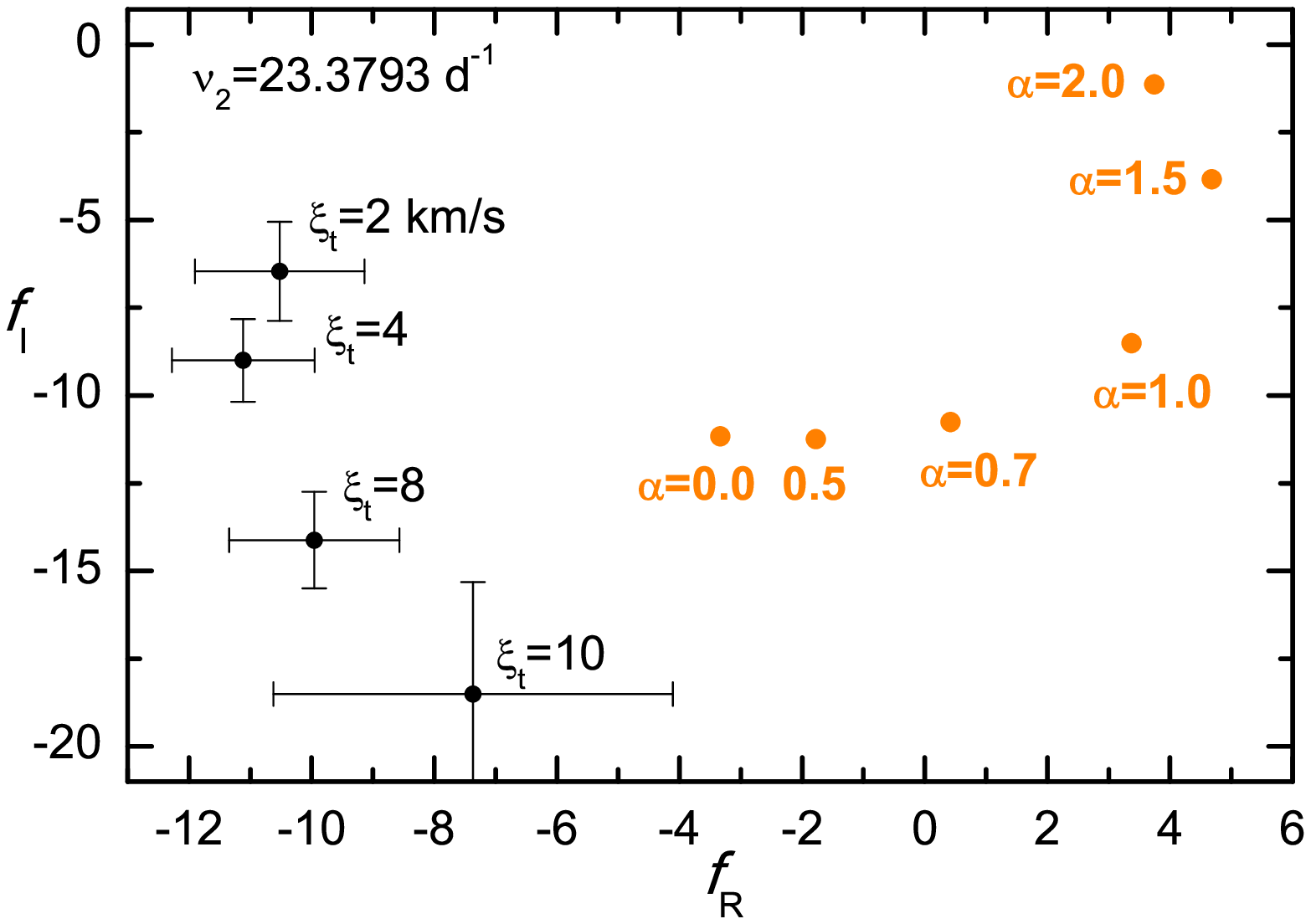}
  \caption{A comparison of the theoretical and empirical values of $f$ on the complex plane for the radial fundamental mode (the left panel)
   and for the first overtone mode (the right panel). The theoretical values were computed for the OPAL seismic model with parameters: $X_0=0.68$, $Z=0.002$,
   $M=1.06 M_\odot,~\log T_{\rm eff}=3.8867,~\log L/L_{\odot}=0.834,~R=1.47 R_\odot$, that rotates with the velocity of about $14~\mathrm{km~s^{-1}}$. The model is marked with a dot on the cyan evolutionary track in Fig.\,3.
   Different values of the mixing length parameter $\alpha_{\rm MLT}$ were considered. The (semi)empirical counterparts were determined from the Str\"omgren photometry and the Vienna model atmospheres assuming different values of the microturbulent velocity $\xi_t$.}
  \label{fig9}
\end{figure*}

In Fig.\,10, we compare the empirical and theoretical values of $f$ for the OPAL seismic model with the chemical $(X_0=0.70,~Z=0.002$ and the parameters:
$M=1.082 M_\odot,~\log T_{\rm eff}=3.8814,~\log L/L_{\odot}=0.819,~R=1.48 R_\odot$. Its rotational velocity is $23~\mathrm{km~s^{-1}}$.
In this case, we got the solution for the empirical values of $f$ determined with the microturbulent velocity $\xi_t=8~\mathrm{km~s^{-1}}$
and the theoretical values of $f$ computed for the mixing length parameter $\alpha_{\rm MLT}\in (0.0,~0.5)$. As before,
we excluded the solution with $\xi_t=10~\mathrm{km~s^{-1}}$ and
$\alpha_{\rm MLT}=0.7$ because of much worse fit of the photometric amplitudes and phases.

Again, for the second frequency, the empirical and theoretical values of $f$ could not be reconciled for any value of $\alpha_{\rm MLT}$ and $\xi_t$.
\begin{figure*}
	\centering
	\includegraphics[width=0.48\textwidth,clip]{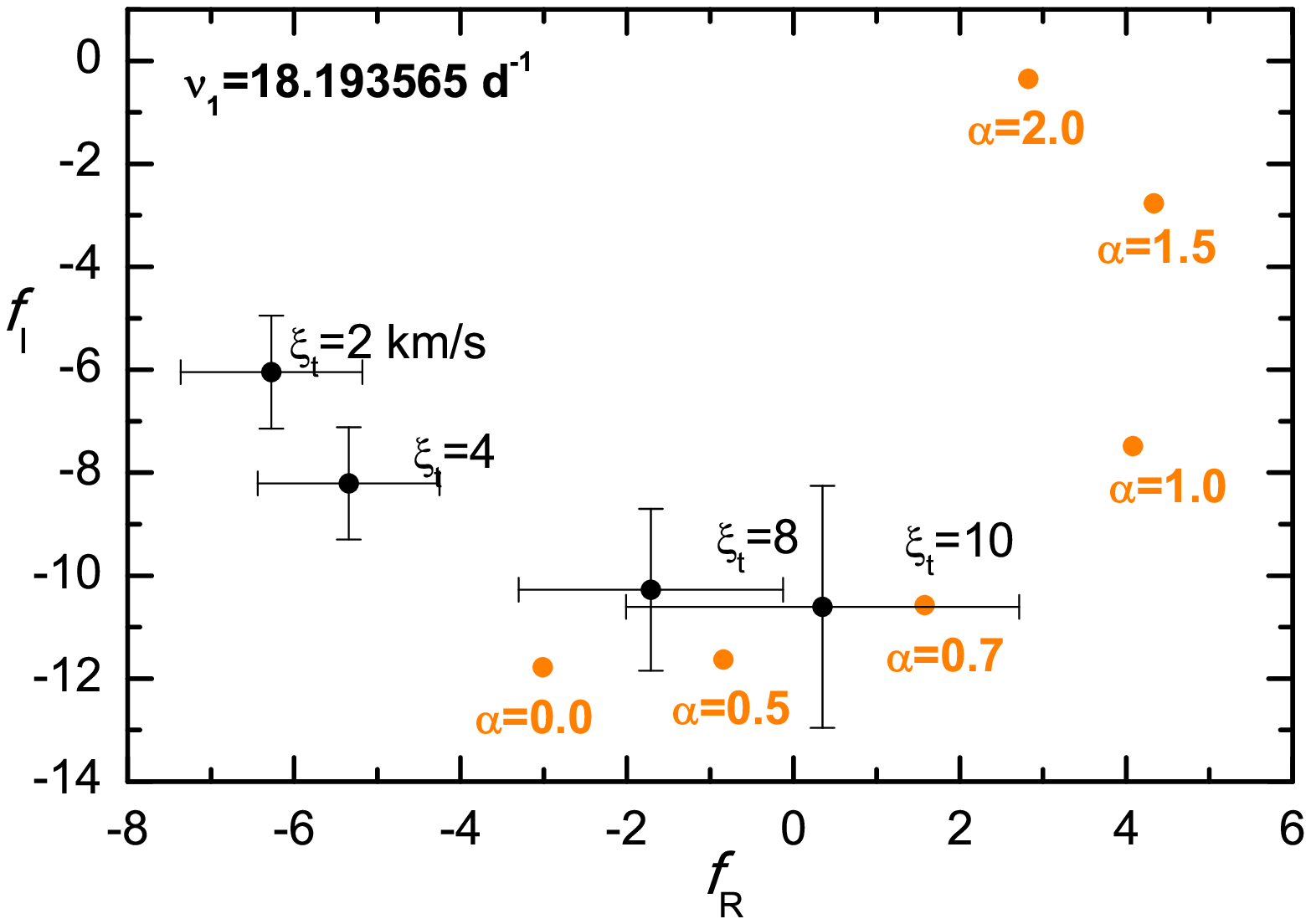}
	\includegraphics[width=0.48\textwidth,clip]{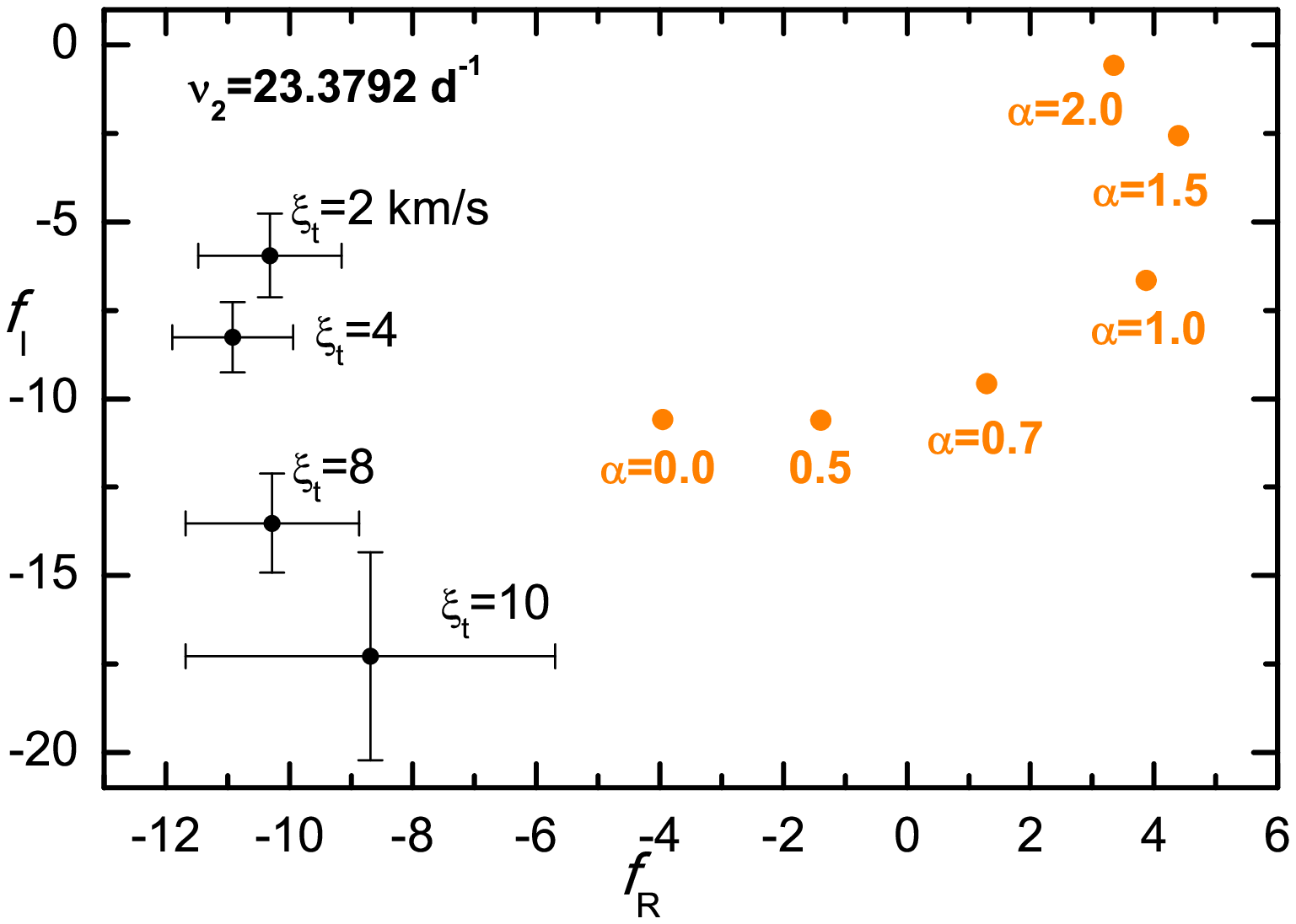}
	\caption{The same comparison as in Fig.\,9 but for the seismic OPAL model with parameters: $X_0=0.70$, $Z=0.002$, $M=1.082 M_\odot$, $\log T_{\rm eff}=3.8814$,
     and $\log L/L_{\odot}= 0.819$ (within the $3\sigma$ error), rotating with the velocity of $23.6~\mathrm{km~s^{-1}}$. The model is marked with a dot on the green evolutionary track in Fig.\,3.}
	\label{fig10}
\end{figure*}



\begin{figure*}
	\centering
	\includegraphics[width=0.48\textwidth,clip]{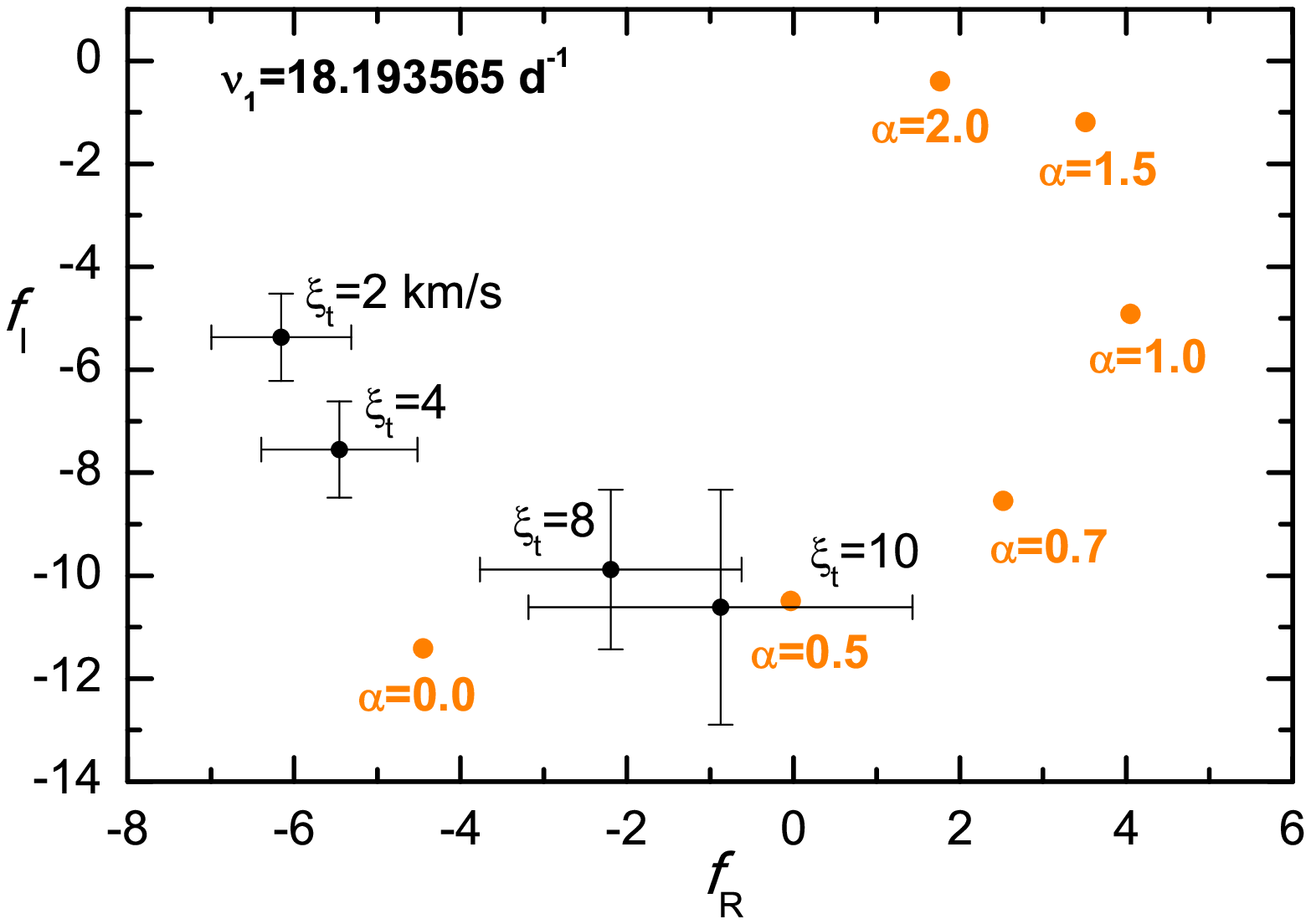}
	\includegraphics[width=0.48\textwidth,clip]{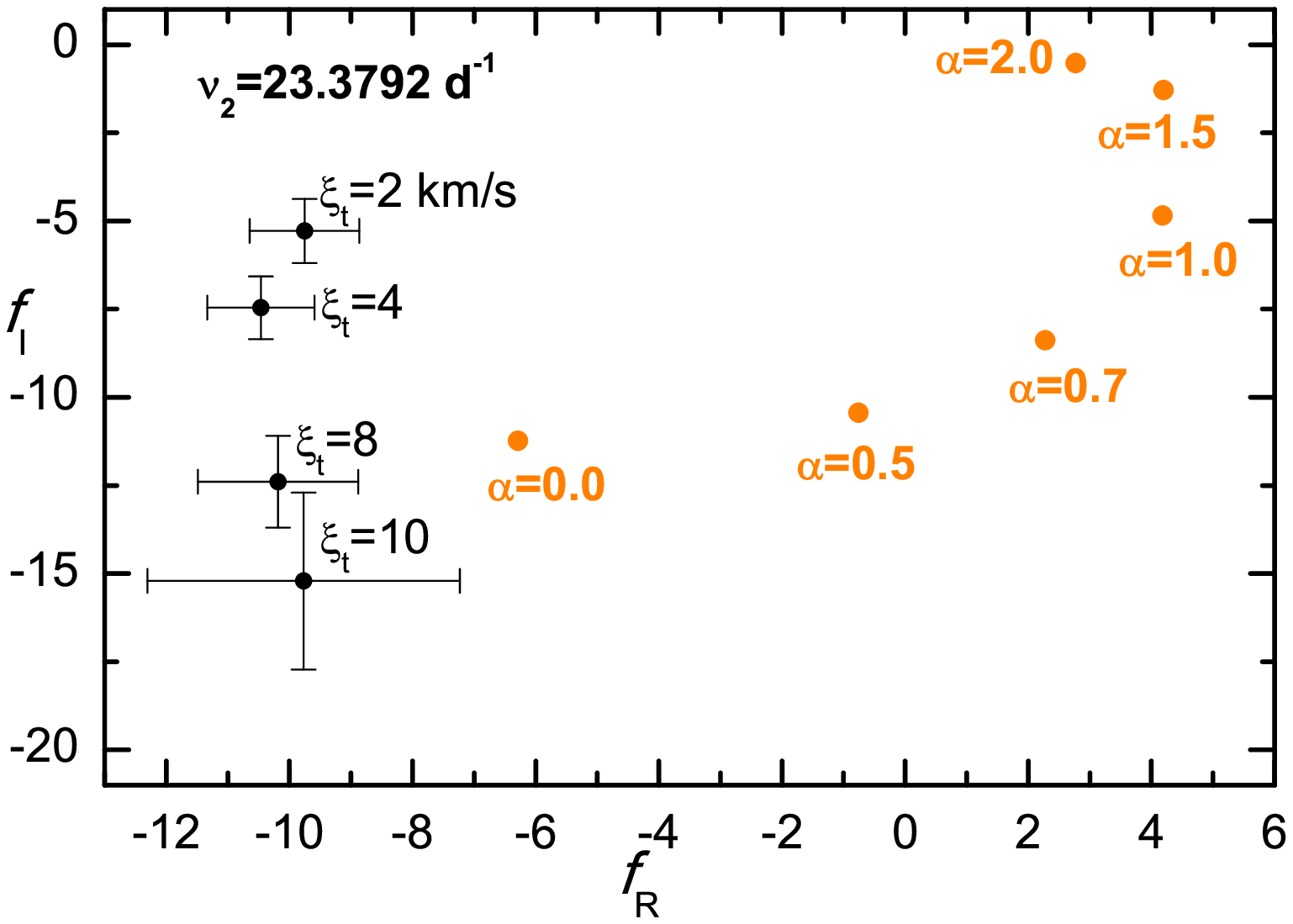}
	\caption{The same comparison as in Fig.\,9 but for the OP seismic model with parameters: $X_0=0.75$, $Z=0.0034$, $M=1.24 M_\odot$, $\log T_{\rm eff}=3.8725$, and $\log L/L_{\odot}= 0.820$ (within the $3\sigma$ error),  $R=1.55~R_\odot$ and the rotational of $17~\mathrm{km~s^{-1}}$.
     The model is marked with a  dot on the dashed evolutionary track in Fig.\,3.}
	\label{fig11}
\end{figure*}

\begin{figure*}
	\centering
	\includegraphics[width=0.48\textwidth,clip]{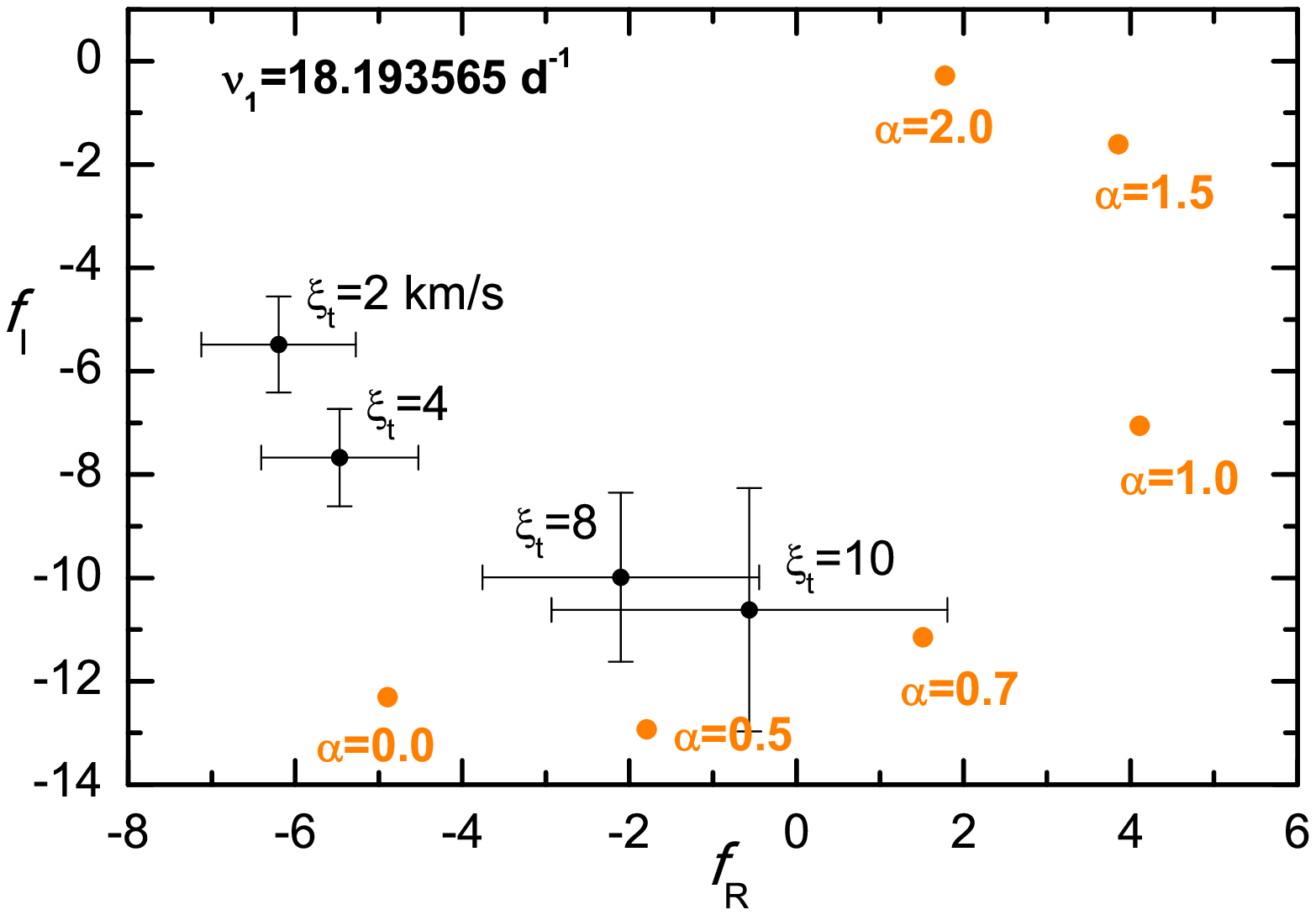}
	\includegraphics[width=0.48\textwidth,clip]{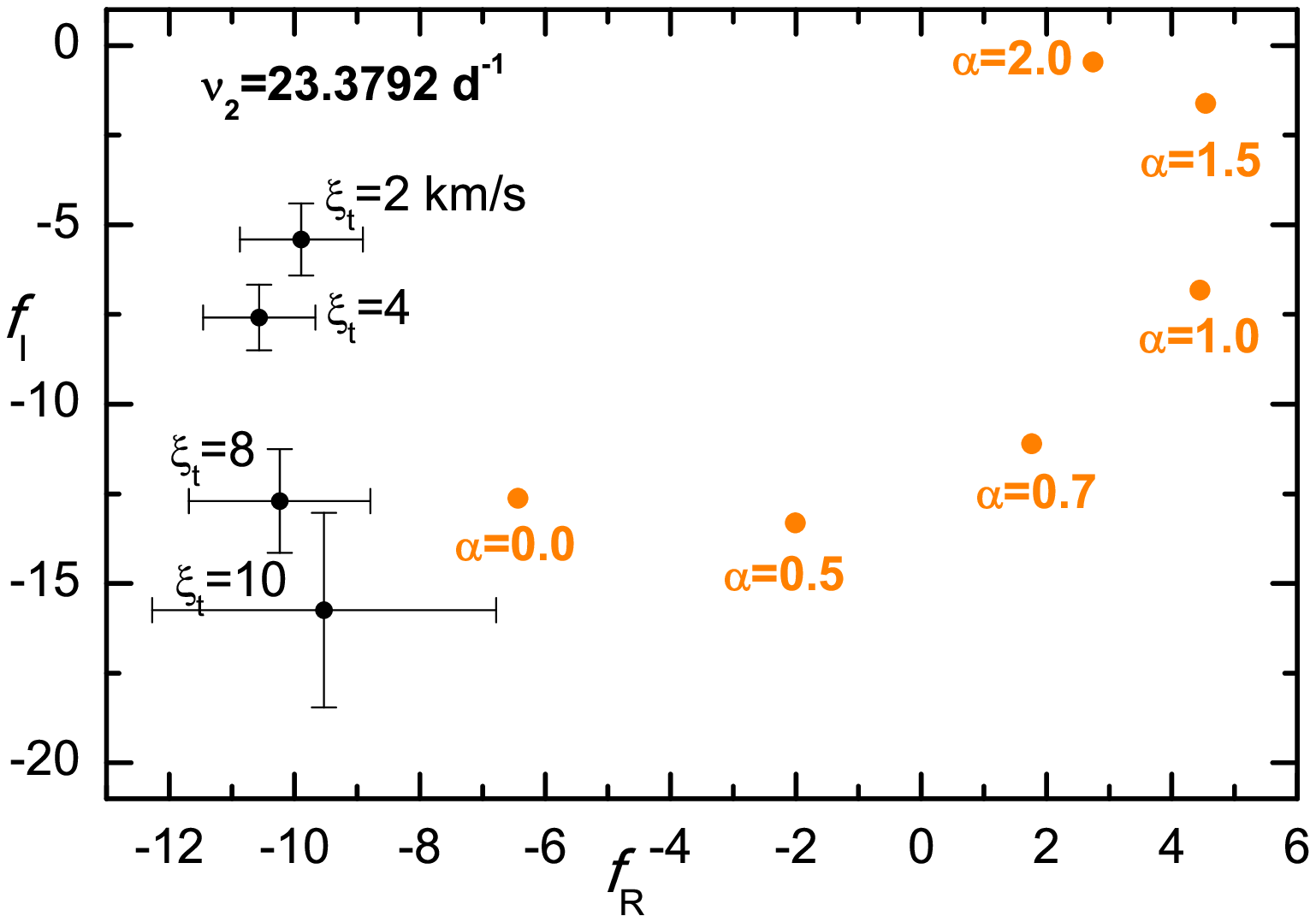}
	\caption{The same comparison as in Fig.\,9 but for the OPLIB seismic model with parameters: $X_0=0.75$, $Z=0.0041$, $M=1.28 M_\odot$, $\log T_{\rm eff}=3.8763$, and $\log L/L_{\odot}= 0.845$,  $R=1.56~R_\odot$ 
	and the rotational velocity of $5.5~\mathrm{km~s^{-1}}$.
     The model is marked with a  dot on the dotted evolutionary track in Fig.\,3.}
	\label{fig12}
\end{figure*}

The next two figures show  comparisons of the theoretical and empirical values of $f$ for seismic models found with
the OP and OPLIB data; Fig.\,11 and Fig.\,12, respectively.
Quite surprisingly, these comparisons are qualitatively similar to
the comparison for the OPAL seismic models (Figs.\,9 and 10), despite of significantly different masses 
and chemical composition.
In the case of the OP seismic model, we can see that the theoretical and empirical values of $f$ of the dominant mode agree
if the mixing length parameter is in the range $\alpha_{\rm MLT}\in (0.0,~0.5)$ and the microturbulent velocity
in the atmosphere is $\xi_t=8~\mathrm{km~s^{-1}}$. For the OPLIB seismic models, the matching for the dominant  mode
was possible for the range of about $\alpha_{\rm MLT}\in (0.5,~0.7)$ and only for  $\xi_t=10~\mathrm{km~s^{-1}}$ .

The first conclusion from Fig.\,9--12 is that, independently of the used opacity data and despite of different parameters of models, constraints on efficiency of the outer-layer convection is very similar. Namely, convection
does not dominate the energy transport in the subphotospheric region
and its efficiency is described by the mixing length parameter of about $\alpha_{\rm MLT}\in (0.0,~0.7)$.

The second conclusion is that, for the range of stellar parameters considered in this work,  
the parameter $f$ is not a diagnostic tool for distinguishing between seismic models calculated
with different opacity data.

\section{Summary}

The goal of this paper was to construct seismic models of the star SX Phoenicis that reproduce all observables
that could be extracted from the available observational data.
Firstly, we made an independent mode identification from the photometric amplitudes and phases
and confirmed pulsations of SX Phe  in the two radial modes. Then, we search for models that fit the observed frequencies of the two radial modes adopting the three commonly
used opacity data for evolutionary computations. The aim was to obtain models with the effective temperature and luminosity
within the adopted observed errors.
Besides, the requirement of instability of both pulsational modes must always be met.

The first result was that only with the OPAL data it was possible to construct such seismic models
having a typical chemical composition for this prototype.
Our best seismic OPAL models have the initial hydrogen abundance in the range $X_0\in(0.67,~0.70)$
and metallicity $Z=0.002$.
Their masses are of about $M=1.05-1.08~\mathrm{M}_\odot$,  the radii: $R=1.47-1.48~R_\odot$ and the age: $2.80-3.07$ Gyr.
All seismic  models  are in the post-main sequence phase and burn hydrogen in the shell.
The best seismic model of SX Phe found by \citet{Petersen1996} with the old OPAL opacities \citep{Iglesias1992} had 
a mass $M=1.0~\mathrm{M}_{\odot}$, metallicity $Z = 0.001$, initial hydrogen
abundance $X_0 = 0.70$ and the age 4.07 Gyr. Thus, it was much older than our best OPAL seismic models.

Then we searched for seismic models using OP and OPLIB opacity data.
It appeared that in these cases, only models computed for significantly higher abundances
of the initial hydrogen and metallicity are able to account for the two radial mode frequencies and have luminosity
within the $3\sigma$ error.
The OP seismic model have a mass $M=1.24~\mathrm{M}_{\odot}$, radius $R=1.55~R_\odot$ and the chemical
composition $X_0=0.75$, $Z=0.0034$. It is after an overall contraction, in the hydrogen shell-burning phase,
and its age is 2.62 Gyr.
The OPLIB seismic model has similar parameters, namely: $M=1.28~\mathrm{M}_{\odot}$,  $R=1.56~R_\odot$,
 $X_0=0.75$, $Z=0.0041$. The model has just finished the overall contraction phase and its age is 2.31 Gyr.
However, we consider the OP and OPLIB  seismic models as less reliable because of their high initial hydrogen abundance $(X_0=0.75)$ and metallicity $(Z=0.0034-0.0041)$. 
Firstly, if SX Phoenicis would be a blue straggler, then it should have rather
enhanced helium abundance \citep[e.g.,][]{McNamara2011, Nemec2017}, hence lower hydrogen abundance.
Secondly, the metallicity of  the OP and OPLIB  seismic models is higher than usually assumed for Population II stars.
However,  we currently do not have firm observables to rule out these models for certain.

In the next step, we used another seismic tool, that is the empirical values of the parameters $\varepsilon$ and $f$
derived from multicolour photometry.  A comparison of the theoretical and empirical values of $f$ for the dominant mode
indicated low to moderately efficient convection, described by the mixing length parameter $\alpha_{\rm MLT}\in (0.0,~0.7)$.
This conclusion is independent of the used opacity data, that is the same values of $\alpha_{\rm MLT}$ were estimated
for seismic models computed with the OPAL, OP and OPLIB tables.
Moreover, this analysis showed that the microturbulent  velocity in the atmosphere amounts 
to about $\xi_{\rm t}\in(4,~8)~\kms$.
Thus, despite a low mass of SX Phe ($M\approx 1.06-1.08~\mathrm{M}_\odot$ for the OPAL seismic models), 
convective transport in its outer layers
is not very efficient. This is because  the radiative gradient is significantly reduced due to much lower opacity
caused by very low metallicity.

The intrinsic mode amplitude, $\varepsilon$, of both frequencies derived from multicolour photometry are consistent with the values
obtained from the radial velocity amplitudes. The relative radius variation for the radial fundamental mode is about 2 per cent
and for the first overtone about 0.3 per cent.

Such comprehensive seismic studies of pulsators like SX Phoenicis are very important for deriving constraints
on convection in the outer layers, because the star is on the border between very efficient and inefficient convection.
More stringent constraints from our complex seismic modelling could be derived for pulsators in double-lined eclipsing binaries.
We plan to make such studies in the near future.

\section*{Acknowledgements}
The work was financially supported by the Polish NCN grant 2018/29/B/ST9/02803.

\section*{Data Availability}
Data on the photometric amplitudes and phases  underlying this article are available in \citet{Rolland1991}.
Our theoretical computations will be shared on reasonable request to the corresponding author.

\bibliographystyle{mnras}
\bibliography{JDD_biblio} 


\bsp	
\label{lastpage}
\end{document}